\documentstyle{article}
\catcode`\@=11
\def\marginnote#1{}

\newcount\hour
\newcount\minute
\newtoks\amorpm
\hour=\time\divide\hour by60
\minute=\time{\multiply\hour by60 \global\advance\minute by-\hour}
\edef\standardtime{{\ifnum\hour<12 \global\amorpm={am}%
        \else\global\amorpm={pm}\advance\hour by-12 \fi
        \ifnum\hour=0 \hour=12 \fi
        \number\hour:\ifnum\minute<10 0\fi\number\minute\the\amorpm}}
\edef\militarytime{\number\hour:\ifnum\minute<10 0\fi\number\minute}

\def\draftlabel#1{{\@bsphack\if@filesw {\let\thepage\relax
   \xdef\@gtempa{\write\@auxout{\string
      \newlabel{#1}{{\@currentlabel}{\thepage}}}}}\@gtempa
   \if@nobreak \ifvmode\nobreak\fi\fi\fi\@esphack}
        \gdef\@eqnlabel{#1}}
\def\@eqnlabel{}
\def\@vacuum{}
\def\draftmarginnote#1{\marginpar{\raggedright\scriptsize\tt#1}}

\def\draft{\oddsidemargin -.5truein
        \def\@oddfoot{\sl preliminary draft \hfil
        \rm\thepage\hfil\sl\today\quad\militarytime}
        \let\@evenfoot\@oddfoot \overfullrule 3pt
        \let\label=\draftlabel
        \let\marginnote=\draftmarginnote
   \def\@eqnnum{(\theequation)\rlap{\kern\marginparsep\tt\@eqnlabel}%
\global\let\@eqnlabel\@vacuum}  }

\begingroup\ifx\undefined\newsymbol \else\def\input#1 {\endgroup}\fi
\input amssym.def \relax
\input amssym
\newfont{\hr}{msbm10}
\newfont{\ams}{msam10}

\def \det{{\rm det}}

\def \log {{\rm log}}

\def \be {begin {equation}}
\def \ee {end{equation}}
\def\tilde{\widetilde}
\def\bar{\overline}
\def\hat{\widehat}
\def\*{\star}
\def\({\left(}		
\def\){\right)}		
\def\[{\left[}		
\def\]{\right]}

\def\frac#1#2{{#1 \over #2}}

\def\2pi{\hbox{$2\pi i$}}

\def\dsl{\raise.15ex\hbox{/}\kern-.57em\partial}
\def\Dsl{\,\raise.15ex\hbox{/}\mkern-.13.5mu D}
\def\be{\beta}

		\def\CF{{\cal F}}

\font\numbers=cmss12
\font\upright=cmu10 scaled\magstep1
\def\stroke{\vrule height8pt width0.4pt depth-0.1pt}
\def\topfleck{\vrule height8pt width0.5pt depth-5.9pt}
\def\botfleck{\vrule height2pt width0.5pt depth0.1pt}
\def\Zmath{\vcenter{\hbox{\numbers\rlap{\rlap{Z}\kern 0.8pt\topfleck}\kern
2.2pt
                   \rlap Z\kern 6pt\botfleck\kern 1pt}}}
\def\Qmath{\vcenter{\hbox{\upright\rlap{\rlap{Q}\kern
                   3.8pt\stroke}\phantom{Q}}}}
\def\Nmath{\vcenter{\hbox{\upright\rlap{I}\kern 1.7pt N}}}
\def\Cmath{\vcenter{\hbox{\upright\rlap{\rlap{C}\kern
                   3.8pt\stroke}\phantom{C}}}}
\def\Rmath{\vcenter{\hbox{\upright\rlap{I}\kern 1.7pt R}}}
\def\Z{\ifmmode\Zmath\else$\Zmath$\fi}
\def\Q{\ifmmode\Qmath\else$\Qmath$\fi}
\def\N{\ifmmode\Nmath\else$\Nmath$\fi}
\def\C{\ifmmode\Cmath\else$\Cmath$\fi}
\def\R{\ifmmode\Rmath\else$\Rmath$\fi}

\newcounter{app}

\def\app{\setcounter{equation}{0}
\def\theequation{\Alph{app}.\arabic{equation}}\par
   \addvspace{4ex}
   \@afterindentfalse
  \secdef\@app\@dapp}

\newcommand\@app{\@startsection {app}{1}{0ex}%
                                   {-3.5ex \@plus -1ex \@minus -.2ex}%
                                   {2.3ex \@plus.2ex}%
                                   {\normalfont\Large\bf}}
\def\@dapp#1{%
{\parindent \z@ \raggedright  \bf #1}\par\nobreak}
\def\l@app#1#2{\ifnum \c@tocdepth >\z@
    \addpenalty\@secpenalty
    \addvspace{1.0em \@plus\p@}%
    \setlength\@tempdima{8em}%
    \begingroup
      \parindent \z@ \rightskip \@pnumwidth
      \parfillskip -\@pnumwidth
      \leavevmode \bfseries
      \advance\leftskip\@tempdima
      \hskip -\leftskip
      #1\nobreak\hfil \nobreak\hb@xt@\@pnumwidth{\hss #2}\par
    \endgroup\fi}

\def\stackreb#1#2{\mathrel{\mathop{#2}\limits_{#1}}}
\def\Tr{{\rm Tr}}
\def\res{{\rm res}}
\def\Bf#1{\mbox{\boldmath $#1$}}

\def\bphi{{\Bf\phi}}

\def\f{1\over}
\def\rank{{\rm rank}}
\def\2{{1\over 2}}
\def\N2{${\cal N}=2$}

\def\be{ \begin{eqnarray} }
\def\ee{ \end{eqnarray} }

\def\bea{\begin{eqnarray}}
\def\eea{\end{eqnarray}}
\def\nn{\nonumber}

\def\beq{\begin{equation}}
\def\eeq{\end{equation}}
\def\ba{\beq\new\begin{array}{c}}
\def\ea{\end{array}\eeq}
\def\be{\ba}
\def\ee{\ea}
\def\stackreb#1#2{\mathrel{\mathop{#2}\limits_{#1}}}
\def\f{1\over}

\def\bz{\bar\zeta}
\def\SUM{\sum_k^{N_c}\bz (a_{N_ck})-\2\sum_\beta^{N_f}\bz (a_{N_c}-m_\beta)} 

\makeatletter
\newdimen\normalarrayskip              
\newdimen\minarrayskip                 
\normalarrayskip\baselineskip
\minarrayskip\jot
\newif\ifold             \oldtrue            \def\new{\oldfalse}
\def\arraymode{\ifold\relax\else\displaystyle\fi} 
\def\eqnumphantom{\phantom{(\theequation)}}     
\def\@arrayskip{\ifold\baselineskip\z@\lineskip\z@
     \else
     \baselineskip\minarrayskip\lineskip2\minarrayskip\fi}
\def\@arrayclassz{\ifcase \@lastchclass \@acolampacol \or
\@ampacol \or \or \or \@addamp \or
   \@acolampacol \or \@firstampfalse \@acol \fi
\edef\@preamble{\@preamble
  \ifcase \@chnum
     \hfil$\relax\arraymode\@sharp$\hfil
     \or $\relax\arraymode\@sharp$\hfil
     \or \hfil$\relax\arraymode\@sharp$\fi}}
\def\@array[#1]#2{\setbox\@arstrutbox=\hbox{\vrule
     height\arraystretch \ht\strutbox
     depth\arraystretch \dp\strutbox
     width\z@}\@mkpream{#2}\edef\@preamble{\halign
\noexpand\@halignto
\bgroup \tabskip\z@ \@arstrut \@preamble \tabskip\z@ \cr}%
\let\@startpbox\@@startpbox \let\@endpbox\@@endpbox
  \if #1t\vtop \else \if#1b\vbox \else \vcenter \fi\fi
  \bgroup \let\par\relax
  \let\@sharp##\let\protect\relax
  \@arrayskip\@preamble}
\def\eqnarray{\stepcounter{equation}%
              \let\@currentlabel=\theequation
              \global\@eqnswtrue
              \global\@eqcnt\z@
              \tabskip\@centering
              \let\\=\@eqncr
              $$%
 \halign to \displaywidth\bgroup
    \eqnumphantom\@eqnsel\hskip\@centering
    $\displaystyle \tabskip\z@ {##}$%
    \global\@eqcnt\@ne \hskip 2\arraycolsep
         $\displaystyle\arraymode{##}$\hfil
    \global\@eqcnt\tw@ \hskip 2\arraycolsep
         $\displaystyle\tabskip\z@{##}$\hfil
         \tabskip\@centering
    &{##}\tabskip\z@\cr}
\def\theequation{\thesection.\arabic{equation}}

\begin{document}
\begin{titlepage}
\setcounter{footnote}0
\begin{center}
\hfill FIAN/TD-15/97\\
\hfill ITEP/TH-63/97\\
\hfill NBI-HE-97-61\\
\vspace{0.4in}
{\LARGE\bf 5d and 6d Supersymmetric Gauge Theories:}\\
\vspace{0.1in}
{\LARGE\bf Prepotentials from Integrable Systems}\\
\bigskip
\bigskip
\bigskip
{\Large A.Marshakov 
\footnote{E-mail address: mars@lpi.ac.ru, andrei@heron.itep.ru, 
marshakov@nbivms.nbi.dk}} 
\\ 
\bigskip 
{\it Niels Bohr Institute, Blegdamsvej 17, 2100 Copenhagen, Denmark\\and\\
Theory Department,  P.N.Lebedev Physics 
Institute, Leninsky prospect 53, Moscow ~117924, Russia\\ 
and ITEP, Moscow 117259, Russia
\footnote{Permanent address}}\\
\bigskip
and\\
\bigskip
{\Large A.Mironov
\footnote{E-mail address:
mironov@lpi.ac.ru, mironov@itep.ru}}\\
\bigskip 
{\it Theory Department,  P.N.Lebedev Physics 
Institute, Leninsky prospect 53, Moscow ~117924, Russia\\ 
and ITEP, Moscow 117259, Russia}
\end{center}
\bigskip \bigskip

\begin{abstract}
We discuss $5d$ and $6d$ supersymmetric gauge theories in the target-space
with compactified directions and
with the matter hypermultiplets in fundamental representations in the 
framework of integrable systems. In particular, we consider the prepotentials
of these theories and derive explicit formulas for their perturbative parts. 
\end{abstract}

\end{titlepage}

\newpage
\section{Introduction}
\setcounter{equation}{0}
\setcounter{footnote}0

In this paper we continue our discussion on the Coulomb branches of gauge
theories, which admit an exact treatment in the low-energy sector \cite{SW}
and are related to integrable systems \cite{GKMMM}-\cite{ggm}.
These are the theories with eight real
supercharges (in accordance with common terminology that means \N2 SUSY in 
four dimensions and ${\cal N}=1$ SUSY in $D=5,6$) with only four noncompact 
directions to have
non-trivial instantonic contribution and, therefore, a non-trivial
prepotential. It means, in particular, that we consider $5d$ theories with the 
target space of topology
$R^4 \times S^1$ and $6d$ theories with the target space of topology
$R^4 \times T^2$. Below we give a detailed analysis of the results for the
prepotentials of these higher dimensional theories. This paper continues our
investigation of the prepotentials started in \cite{wdvv} and, in
particular, gives some support to the integrable structures of higher
dimensional theories proposed in \cite{MMaM,ggm}.

We consider here the SUSY theories with massive hypermultiplets
in fundamental representations. Therefore, in accordance with
\cite{xxx,xyz,ggm}, the corresponding
integrable structures are given by (classical) spin chains:
the $XXZ$ chain in $5d$ and $XYZ$ (Sklyanin) chain in $6d$. 
Below we concentrate on the issues related to the
prepotentials and associated analytic structures on Riemann surfaces:
differentials, their periods etc. In particular, we calculate the perturbative
prepotentials which coincide with the expected from general considerations
of quantum field theory. It gives some
support to the identification of the integrable spin-chains with the SUSY 
QCD in higher dimensions. 

An integrable system, i.e. a spectral curve and a generating 1-differential
$dS$ \cite{DKN}, allows to define the prepotential in the framework of the 
Seiberg-Witten (SW) anzatz \cite{SW}, where it can be identified with the
(logarithm of) tau-function of the Whitham hierarchy \cite{GKMMM}.
In sect.2 we remind this procedure for the standard case of
$4d$ \N2 QCD. In sect.3 we  extend it to the $5d$ case, in particular,
we advocate there the most effective method of calculating the prepotentials
based on residue formulas first studied in the framework of the SW theory in
connection with the associativity equations \cite{wdvv1,wdvv}.
The procedure becomes especially explicit when computing the perturbative
prepotentials.
In sect.4 we discuss $6d$ theories described by the $XYZ$ spin
chain and sect.5 contains some discussion. Appendix contains
some definitions and formulas on elliptic functions we use
throughout sect.~4.

\section{$4d$ supersymmetric theories: the Seiberg-Witten anzatz}
\setcounter{equation}{0}
First, as a warm-up example we consider the integrable system
corresponding to the $4d$ SUSY QCD
and explain how the data obtained from integrable theory
can be associated with the objects in nonperturbative SUSY theories,
in the context of the Seiberg-Witten anzatz \cite{SW}.
Our presentation follows the line worked out in \cite{Mgur,xxx,xyz,wdvv,ggm} 
and some details omitted below can be possibly found there.

\subsection{$4d$ theories and $XXX$ chain}

Let us remind, first, the definition of the $sl(2)$ (integrable) $XXX$ spin 
chain. The most convenient formulation is based on the $2\times 2$ Lax 
matrix (see for example \cite{FT}), which for the $sl(2)$ $XXX$ chain is
\be\label{laxmatr}
L(\lambda) = \lambda \cdot {\bf 1} + \sum_{a=1}^3 S_a\cdot\sigma^a.
\ee
and the Poisson brackets on the space of the dynamical variables $S_a$, 
$a=1,2,3$ in the $2\times 2$ formalism are implied by {\em quadratic} r-matrix 
relations \cite{Skl}
\be\label{quadr-r}
\left\{L(\lambda)\stackrel{\otimes}{,}L(\lambda')\right\} =
\left[ r(\lambda-\lambda'),\ L(\lambda)\otimes L(\lambda')\right],
\ee
with the rational $r$-matrix 
\be\label{rat-r}
r(\lambda) = \frac{1}{\lambda}\sum_{a=1}^3 \sigma^a\otimes \sigma^a.
\ee
Following standard procedure \cite{xxx}, one should consider
the {\em chain} of the Lax operators (\ref{laxmatr}) associated to
each site of the chain and commuting with each other at different sites.
The linear problem in the spin chain \cite{FT} can be defined by
\be\label{lprob}
L_i(\lambda)\Psi_i(\lambda)=\Psi_{i+1}(\lambda)
\ee
where $\Psi_i(\lambda)$ is the two-component Baker-Akhiezer function.
One can introduce the transfer matrix acting from $i$-th site to $i+n$-th
taking product of the Lax matrices (\ref{Tmat}).
The crucial property of the relation (\ref{quadr-r}) is that it
is multiplicative, i.e. any transfer matrix, being product of the form
\be\label{Tmat}
T_n(\lambda)\equiv L_n(\lambda)\ldots L_1(\lambda)
\\
T_n(\lambda)\Psi_i(\lambda)=\Psi_{i+n}(\lambda)
\ee
satisfies the same relation
\be\label{Tbr}
\left\{T_n(\lambda)\stackrel{\otimes}{,}T_n(\lambda')\right\} =
\left[ r(\lambda-\lambda'),\
T_n(\lambda)\otimes T_n(\lambda')\right],
\ee
provided all $L_i$ in product (\ref{Tmat})
are independent, $\{L_i,L_j\} = 0$ for $i\neq j$.
The relations (\ref{quadr-r}) are also true for the {\em inhomogeneous} 
$T$-matrix ($\lambda\to\lambda + \lambda_i$ at $i$-th site; 
in the applications to \N2 SUSY gauge theories the chain has to be taken
of the length $N_c$ \cite{xxx})
\be\label{T-matrix}
T_{N_c}(\lambda) = \prod_{N_c\ge i\ge 1}^{\curvearrowleft}
L_i(\lambda-\lambda_i)
\ee
still satisfying (\ref{Tbr}) with the same $r$-matrix (\ref{rat-r}).

In the $sl(2)$ $XXX$ case the $r$-matrix relations (\ref{quadr-r}) are
equivalent just to the common $sl(2)$ commutation relation (written
in terms of the Poisson brackets)
\be\label{Scomrel}
\{S_a,S_b\} = i\epsilon_{abc} S_c,
\ee
i.e. $\{S_a\}$ plays the role of angular momentum (``classical spin'') 
variables giving the name ``spin-chains'' to the whole class of systems.
The algebra (\ref{Scomrel}) has an obvious Casimir operator
(an invariant, which Poisson commutes with all the generators $S_a$),
\be\label{Cas}
K^2 =  \sum_{a=1}^3 S_aS_a,
\ee
so that
\be\label{detTxxx}
\det_{2\times 2} L(\lambda) = \lambda^2 - K^2,
\nn \\
\det_{2\times 2} T_{N_c}(\lambda) = \prod_{i=1}^{N_c}
\det_{2\times 2} L_i(\lambda-\lambda_i) =
\prod_{i=1}^{N_c} \left((\lambda - \lambda _i)^2 - K_i^2\right) = \nn \\
= \prod_{i=1}^{N_c}(\lambda - m_i^+)(\lambda - m_i^-)
= Q_{2N_c}(\lambda),
\ee
where we assumed that the values of spin $K$ can be different at
different nodes of the chain, and
\footnote{
Eq.(\ref{mpm}) implies that the limit of vanishing masses, all
$m_i^\pm = 0$, is associated with the {\it homogeneous} chain
(all $\lambda_i = 0$) and vanishing spins at each site (all $K_i = 0$).}
\be
m_i^{\pm} = \lambda_i \pm K_i.
\label{mpm}
\ee
While the determinant of monodromy matrix (\ref{detTxxx})
depends on dynamical variables
only through the Casimir functions $K_i$ of the Poisson algebra, the 
dependence of
\be\label{trP}
P_{N_c}(\lambda) =
\frac{1}{2}\Tr_{2\times 2}T_{N_c}(\lambda) =\prod_{i=1}^{N_c}
\left(\lambda-\lambda_i\right)
\ee
upon dynamical variables is less trivial.
Still, as usual for integrable systems
\be
\{ P_{N_c}(\lambda ),P_{N_c}(\lambda ')\} = {1\over 4}
\{ \Tr T_{N_c}(\lambda ),\Tr T_{N_c}(\lambda ')\} = {1\over 4}
\Tr\left[ r(\lambda-\lambda'),\ T_{N_c}(\lambda)\otimes T_{N_c}
(\lambda')\right] = 0
\ee 
i.e.  the polynomial $P_{N_c}(\lambda)$ (\ref{trP}) depends
on $S_a^{(i)}$ only through the Hamiltonians of spin chain (Poisson-commuting 
with {\it each other}).

The (quasi)periodic boundary conditions imposed to the Baker-Akhiezer function 
\be\label{pbc}
\Psi_{i+N_c}(\lambda) = w\Psi_{i}(\lambda)
\ee
give rise to the
spectral curve equation for the periodic spin chain
\be\label{fsc-SCh}
\det\left(T_{N_c}(\lambda) -  w\right) = 0,
\ee
In the particular case of $sl(2)$ $XXX$ spin chains, the spectral
equation acquires the form
\footnote{In general case of $sl(p)$ chains the spectral
equation is of the $p$-th order in $w$, see \cite{xxx,ggm} and below.}
\be\label{fsc-sc1}
w + \frac{Q_{N_f}(\lambda)}{w} = 2P_{N_c}(\lambda )
\ee
where the number of matter multiplets $N_f\le 2N_c$ depending on a particular
``degeneracy'' of the full chain and, as follows from (\ref{laxmatr}), 
(\ref{trP}) $\sum\lambda_i=0$.

It is relatively easy to get the exact expressions for the Hamiltonians
(the explicit examples of monodromy matrices and Hamiltonians
for $N_c = 2$ and $3$ can be found in \cite{xxx}). 
The Hamiltonians depend non-\-trivially
on the $\lambda_i$-parameters (in\-homo\-geneities of the chain)
and the coefficients
in the spectral equation (\ref{fsc-SCh}) depend only on the
Hamiltonians and symmetric functions of the mass-parameters (\ref{mpm}),
i.e. the dependence of $\{\lambda_i\}$ and $\{K_i\}$ is rather special.
This property is crucial for identification of the $m$-parameters
with the masses of the matter supermultiplets in the ${\cal N}=2$ SQCD.

\subsection{Spectral curves and prepotentials}

Let us turn now to the $4d$ SW anzatz \cite{SW}, which
can be formulated in the following way.
The \N2 SUSY vector multiplet has necessarily (complex) scalars 
with the potential $V(\bphi )=\Tr [\bphi, \bphi ^{\dagger}]^2$ whose 
minima (after factorization over the 
gauge group) correspond to the diagonal ($[\bphi, \bphi ^{\dagger}] = 0$), 
constant and (in the theory with $SU(N_c)$ gauge group) traceless matrices. 
Their invariants 
\be\label{polyn} 
\det (\lambda - \bphi) \equiv P_{N_c}(\lambda ) = \sum_{k=0}^{N_c} 
s_{N_c-k}\lambda ^k 
\ee 
(the total number of algebraically independent ones is 
$\rank\ SU(N_c) = N_c -1$) parametrize the moduli space of the (Coulomb
branch of the) theory. 
Due to the Higgs effect the off-diagonal part of the gauge field 
${\bf A}_{\mu}$ becomes massive, since 
\be\label{comm} 
[\bphi , {\bf A}_{\mu}]_{ij} = (\phi_i-\phi_j){\bf A}_{\mu}^{ij} 
\ee 
while the diagonal part, as it follows from (\ref{comm}) remains massless, 
i.e. the gauge group $G = SU(N_c)$ breaks down to 
$U(1)^{\rank G} = U(1)^{N_c -1}$. 

The effective abelian $U(1)^{N_c -1}$ theory is formulated in terms of a 
(finite-dimensional) integrable system: the spectral curve $\Sigma $ 
defined over the genus-dimensional subspace of the full moduli space, e.g. 
\be\label{suncu} 
{\cal P}(\lambda ,w) = 2P_{N_c}(\lambda) - w - {1\over w} = 0
\ee 
for the pure $SU(N_c)$ gauge theory  and the generating 
differential 
\be\label{dS} 
dS = \lambda{dw\over w} 
\ee 
whose basic property is that its derivatives over $N_c-1$ moduli give rise 
to holomorphic differentials. From the point of view of integrable
system, the spectral curve express the condition of the common 
spectrum of two operators, usually the Lax operator ${\cal L}$ and some
auxiliary operator, which was taken above to be the shift operator $T_{N_c}$
(\ref{Tmat}). From practical point of view one can work in the basis of
eigenvalues of one of the operators: in the formalism presented above we
have chosen the basis of the Lax eigenfunctions -- the wave functions which
explicitly depend on the Lax eigenvalue $\lambda $, then another operator
acts in this basis as a $\lambda$-dependent matrix. 
As a result, the full spectral curve $\Sigma$ arises as a
ramified covering over some {\em bare} spectral curve:
\be
{\cal P}(\lambda; z) = 0
\ee
where $\lambda $ or $z$ ($z=\log w$ in (\ref{suncu})) can be considered as a 
co-ordinate on bare spectral
curve $\Sigma_0$. In the case of the gauge group  $G=SU(N_c)$, 
the function ${\cal P}$, caused by (\ref{polyn}), is a polynomial of degree 
$N_c$ in $\lambda$. If one adds the fundamental matter hypermultiplets 
\cite{fumat} one should also complete the set of moduli by their masses
$m_{\alpha}$ ($\alpha = 1,\dots,N_f$) and modify the form of the curve 
(\ref{suncu}) to (\ref{fsc-sc1}). Then, natural  generating differential is
\be\label{dSqcd'}
dS=\lambda{dw\over w}
\ee
There is, however, another generating differential usually choosen
in literature \cite{SW,xxx,fumat}, which is expressed through the variable
$W \equiv {w\over\sqrt{Q_{N_f}(\lambda)}}$:
\be\label{dSqcd}
dS=\lambda{dW\over W}
\ee
This variable is naturally associated with the curve written in some
more symmetric form\footnote{In terms of brane
picture it depends on the exact form of embedding of a spectral curve into
(eleven-dimensional) target space (see, for example, \cite{W,MMaM,ggm,theya}
for more detailed discussion of this point).}
\be\label{sqcd} 
W + {1\over W} = {2P_{N_c}(\lambda)\over 
\sqrt{Q_{N_f}(\lambda)}} 
\ee
The both choices of the differentials lead to the same results \cite{wdvv}.
In this paper, we use the differential (\ref{sqcd}) as better suited for 
perturbative calculations.

The coefficients  $s_I$ of the function ${\cal P}$ parametrize the 
(subspace of) moduli 
space of complex structures ${\cal M}_{\Sigma}$ of the curve $\Sigma$. The 
Hamiltonians or action variables
(integrals of motion) are specific co-ordinates on moduli space. From
the four-dimensional point of view, the co-ordinates  $s_I$ include $s_i$ --
(the Schur polynomials of) the adjoint-scalar expectation values $h_k =
\frac{1}{k}\langle\Tr \bphi^k\rangle$ of the vector ${\cal N}=2$
supermultiplet, as well as $s_\alpha = m_\alpha$ -- the masses of the
hypermultiplets. One associates with each handle of $\Sigma$ a gauge
modulus and with each puncture -- a massive hypermultiplet with mass 
given by residue of $dS$ in the puncture.

The generating 1-form $dS \cong \lambda dz$ is meromorphic on
$\Sigma$ (hereafter the equality modulo total derivatives
is denoted by ``$\cong$''). The prepotential is defined in terms of the
cohomological class of $dS$:
\be
a_I = \oint_{A_I} dS, \ \ \ \ \ \
\frac{\partial \CF}{\partial a_I} = \int_{B_I} dS \nn \\
A_I \circ B_J = \delta_{IJ}.
\label{defprep}
\ee
The cycles $A_I$ include the $A_i$'s wrapping around the handles
of $\Sigma$ and $A_\alpha$'s, going around the singularities
of $dS$.
The conjugate contours $B_I$ include the cycles $B_i$ and the
{\it non-closed} contours $B_\alpha$, ending at the singularities
of $dS$ (see \cite{wdvv} for details).
The integrals $\int_{B_\alpha} dS$ are actually divergent, but
the coefficient of divergent part is equal to residue of $dS$
at particular singularity, i.e. to $a_\alpha$. Thus, the divergent
contribution to the prepotential is quadratic in $a_\alpha$, while
the prepotential is normally defined {\it modulo} quadratic combination
of its arguments (which just fixes the bare coupling constant). In
particular models $\oint_{A,B} dS$ for some pairs of conjugate contours are
identically zero on entire ${\cal M}$: such pairs are not included into the
set of indices $\{I\}$.

Note that the data $(\Sigma, dS)$ with such properties
are {\em exactly} the definition of the integrable system in the sense of
\cite{DKN} (see also \cite{m} and references therein for details). The period
matrix of $\Sigma$: $T_{ij}(a_i)={\partial^2 \CF\over
\partial a_i\partial a_j}$
as a function of the action variables $a_i$
gives the set of coupling constants in the effective abelian
$U(1)^{N_c-1}$ theory while action variables themselves are identified
with the masses of the BPS states $M^2 \sim |{\bf na} + {\bf ma}_D|^2$ with
the $({\bf n},{\bf m})$ "electric" and "magnetic" charges.
As an output of all these definitions, one can derive the
following residue formula \cite{wdvv1,wdvv} for the derivatives of 
prepotential
\be 
\frac{\partial^3 \CF}{\partial a_I\partial a_J\partial a_K} =
\frac{\partial T_{IJ}}{\partial a_K} =
\stackreb{dz  = 0}{\res}
\frac{d\omega_Id\omega_Jd\omega_K}{dz  d\lambda}
\label{resfor}
\ee
where $d\omega_I$ are the canonical holomorphic 1-differentials
satisfying $\oint_{A_I} d\omega_J = \delta_{IJ}$. Their basic
property $\oint_{B_I} d\omega_J = T_{IJ}$ can be obtained by
differentiating the meromorphic differential $dS$ w.r.t. to the moduli $a_I$.

This formula can be effectively used for the calculations of the
prepotentials term by term, in particular, for the perturbative calculations.

\subsection{Perturbative prepotential}

Now, following \cite{wdvv} we discuss the simplest case of degenerate curves 
corresponding to the perturbative formulas. In the perturbative limit 
(${\Lambda_{QCD}\over\langle\bphi\rangle}\rightarrow 0$) 
the second term in (\ref{sqcd}) vanishes, the curve acquires the form 
\be\label{pertcurv} 
W = 2\frac{P_{N_c}(\lambda)}{\sqrt{Q_{N_f}(\lambda)}} 
\ee 
(a rational curve with punctures
\footnote{These punctures emerge as a 
degeneration of the handles of the hyperelliptic surface so that the 
$A$-cycles encircle the punctures.}) and the generating differential 
(\ref{dSqcd}) turns into
\be\label{pertdS} 
dS^{(4)}_{\rm pert} = \lambda d\log\left(\frac{P_{N_c}(\lambda)}
{\sqrt{Q_{N_f}(\lambda)}}\right)
\ee 
We remind, first, the simplest case of the pure gauge $SU(N_c)$ Yang-Mills 
theory (\ref{suncu}), (\ref{dS}), i.e. $Q(\lambda)=1$. 
Now the set of the $N_c-1$ independent 
canonical holomorphic differentials is
\be\label{difff} 
d\omega_i=\left({1\over\lambda-\lambda 
_i}-{1\over\lambda -\lambda_{N_c}}\right)d\lambda= 
{\lambda_{iN_c}d\lambda\over 
(\lambda -\lambda_i)(\lambda -\lambda_{N_c})},\ \ \ i=1,...,N_c-1, 
\ \ \ \lambda_{ij}\equiv \lambda_i-\lambda_j 
\ee 
and the $A$-periods $a_i = \lambda_i$ (the independent ones are, say, with 
$i=1,\dots,N_c-1$) coincide with the roots of polynomial $P_{N_c}(\lambda)$ 
(\ref{trP}).
The prepotential can be now computed via residue formulas \cite{wdvv} 
\be 
\CF_{ijk}=\stackreb{d\log P_{N_c}=0}{\res} {d\omega_id\omega_jd\omega_k\over 
d\log P_{N_c}(\lambda)d\lambda} 
\ee 
which can be explicitly calculated. The only technical trick is that it
is easier to compute the residues at the poles of $d\omega$'s instead of 
zeroes of $d\log P$. This can be done immediately since there are no 
contributions from the infinity $\lambda=\infty$. 
The final results have the following form 
\be  
\CF_{iii}=\sum_{k\ne i}{1\over \lambda_{ik}}+{6\over \lambda_{iN_c}}+ 
\sum_{k\ne N_c}{1\over \lambda_{kN_c}}, 
\\ 
\CF_{iij}= {3\over \lambda_{iN_c}}+{2\over \lambda_{jN_c}}+\sum_{k\ne i,j,N_c} 
{1\over \lambda_{kN_c}}-{\lambda_{jN_c}\over \lambda_{iN_c}\lambda_{ij}}, 
\ \ \ i\ne j,\\ 
\CF_{ijk}=2\sum_{l\ne N_c}{1\over \lambda_{lN_c}}-\sum_{l\ne i,j,k,N_c} 
{1\over \lambda_{lN_c}}, 
\ \ \ i\ne j\ne k; 
\ee
giving rise to the prepotential 
formula 
\be\label{FA}
{\cal F} = {1\over 4}\sum _{ij}f^{(4)}(\lambda_{ij})
\\
f^{(4)}(x) = x^2\log x^2
\ee 
Now let us turn back to the case of $4d$ \N2 QCD with massive hypermultiplets 
Then, there 
arise additional differentials corresponding to the derivatives of $dS$ 
(\ref{pertdS}) with respect to masses. They are of the form 
\be\label{massdiff} 
d\omega_{\alpha}=-{1\over 2}{d\lambda\over \lambda -\lambda_{\alpha}} 
\ee 
It is again straightforward 
to use the residue formula\footnote{Let us note that, despite 
the differentials 
$d\omega_{\alpha}$ have the pole at infinity, this does not contribute 
into the residue formula because of the quadratic pole of $d\lambda$ in the 
denominator.} 
\be\label{residueQ} 
\CF_{IJK}=\stackreb{d\log {P\over\sqrt{Q}}=0}{\res} 
{d\omega_Id\omega_Jd\omega_K\over d\log 
{P(\lambda)\over\sqrt{Q(\lambda)}}d\lambda}, 
\ \ \ \ \ 
\left\{I,J,K,\dots\right\}=\left\{i,j,k,\ldots|\alpha,\beta, 
\gamma,\ldots\right\} 
\ee 
and obtain 
\be\label{perth} 
\CF_{iii}=\sum_{k\ne i}{1\over \lambda_{ik}}+{6\over 
\lambda_{iN_c}}+ \sum_{k\ne N_c}{1\over 
\lambda_{kN_c}}-\2\sum_{\alpha}{\lambda_{iN_c}\over 
\lambda_{i{\alpha}}\lambda_{{\alpha}N_c}}, 
\\ 
\CF_{iij}= {3\over \lambda_{iN_c}}+{2\over \lambda_{jN_c}}+\sum_{k\ne i,j,N_c} 
{1\over \lambda_{kN_c}}- 
{\lambda_{jN_c}\over \lambda_{iN_c}\lambda_{ij}}-\2\sum_{\alpha}{1\over 
\lambda_{{\alpha}N_c}}, 
\ \ \ i\ne j,\\ 
\CF_{ijk}=2\sum_{l\ne N_c}{1\over \lambda_{lN_c}}-\sum_{l\ne i,j,k,N_c} 
{1\over \lambda_{lN_c}}-\2\sum_{\alpha}{1\over\lambda_{{\alpha}N_c}}, 
\ \ \ i\ne j\ne k;\\ 
\CF_{ii{\alpha}}= \2\left({\f\lambda_{i{\alpha}}}-{\f\lambda_{{\alpha}N_c}} 
\right);\\ 
\CF_{ij{\alpha}}=-\2{\f\lambda_{{\alpha}N_c}},\ \ \ i\ne j;\\ 
\CF_{i{\alpha}{\alpha}}=-\2{\lambda_{iN_c}\over\lambda_{i{\alpha}} 
\lambda_{{\alpha}N_c}};\\ 
\CF_{{\alpha}{\alpha}{\alpha}}= 
\2\sum_i{\f\lambda_{{\alpha}i}}+{1\over 4}\sum_{\beta} 
{\f\lambda_{{\alpha}\beta}};\\ 
\CF_{{\alpha}{\alpha}\beta}= 
-{1\over 4}{\f \lambda_{{\alpha}\beta}},\ \ \ {\alpha}\ne \beta;\\ 
\CF_{i{\alpha}\beta}=\CF_{{\alpha}\beta\gamma}=0 
\ee 
These formulas immediately lead to the prepotential 
\be\label{f4d}
\CF={1\over 4}\sum_{i,j}f^{(4)}(a_{ij})-{1\over 4}\sum_{i,\alpha}
f^{(4)}(a_i-m_{\alpha})+{1\over 16}\sum_{\alpha,\beta}
f^{(4)}(m_{\alpha}-m_{\beta})
\ee
upon the 
identification $\lambda_i=a_i$ and $\lambda_{\alpha}=m_{\alpha}$.

\section{$XXZ$ spin chain and $5d$ theories}
\setcounter{equation}{0}
Now we are going to describe the generic integrable system behind 
the $5d$ theories --
the inhomogeneous $XXZ$ spin chain, and reproduce from the integrable
system the main ingredients of the $5d$ SW theory:
the spectral curve of and generating 1-differential $dS$. 
As in $4d$ case it allows to
construct explicitly the holomorphic differentials on spectral curve and 
calculate the perturbative part of $5d$ prepotential.

\subsection{$XXZ$ spin chain}

In $2\times 2$ formalism the generalization of $4d$ formulas is 
straightforward:
the Lax matrix for the $XXZ$ ($sl(2)$) spin magnetic is defined on a cylinder
and has the form \cite{FT}
\be\label{l-gen}
L(\mu)\,=\,\left(
\begin{array}{cc}
\mu e^{S_0}-\mu^{-1}e^{-S_0} & 2S_-\\
2S_+ & \mu e^{-S_0}-\mu^{-1}e^{S_0}
\end{array}\right)
\ee
In fact, from the point of view of integrable systems more natural
spectral parameter is $\xi = \log\mu$ so that the matrix elements of the
Lax operator (\ref{l-gen}) are trigonometric functions of $\xi$.
The Poisson bracket of the Lax operator (\ref{l-gen})
\be\label{quadr-r5}
\left\{L(\mu)\stackrel{\otimes}{,}L(\mu')\right\} =
\left[ r(\mu-\mu'),\ L(\mu)\otimes L(\mu')\right]
\ee
is defined by the trigonometric $r$-matrix
\be\label{trigrmatrix}
r(\xi)={i\over\sinh\pi\xi}
\left(\sigma_1\otimes\sigma_1+\sigma_2\otimes\sigma_2+
\cosh\pi\xi\sigma_3\otimes\sigma_3\right)
\ee
so that (\ref{quadr-r5})
gives rise to the following Poisson algebra of $S_i$'s:
\be\label{pois}
\{S_+,S_0\}=\pm S_{\pm};\ \ \ \{S_+,S_-\}=\sinh 2S_0
\ee
The second Casimir function is now
\be\label{Casimir}
K=\cosh 2S_0+2S_+S_-
\ee
The non-linear relations (\ref{pois}) remind
the commutation relations of {\em quantum} algebra $U_q(sl(2))$,
this is a generalization the fact that the
$XXX$ Poisson brackets (\ref{Scomrel}) reproduce the
{\em classical} algebra $sl(2)$. The value of deformation parameter
$q$ is inessential in (\ref{pois}) and can be put equal to unity; 
in fact, it is proportional to
the radius of the target space circle $R_5$ in corresponding $5d$ SUSY
theory and can be easily restored replacing each generator
$S_i$ by $R_5S_i$. Hereafter, we omit $R_5$ from all the formulas except for
some special cases below.

Again as in $4d$ case one should consider
the chain with $N_c$ sites, now with the Lax operators (\ref{l-gen}) 
associated to each site 
\be\label{lproblem}
L_i(\mu)\Psi_i(\mu)=\Psi_{i+1}(\mu)
\ee
and commuting with each other for $i\neq j$; introduce the inhomogeneities 
$\xi_i$
which depend on the site of chain replacing $\xi\to\xi-\xi_i$
and impose periodic boundary conditions.
As before the transfer matrix acting from $i$-th to $i+N_c$-th site 
is the product of the Lax matrices (\ref{Tmat}) 
\be
T_{N_c}(\mu) = \prod_{N_c\ge i\ge 1}^{\curvearrowleft} 
L_i\left({\mu\over\mu_i}\right)
\ee
and
the (quasi)periodic boundary conditions
\be\label{pbc5}
\Psi_{i+N_c}(\mu)=-w\Psi_{i}(\mu)
\ee
provide the spectral curve equation now for the $XXZ$ spin chain
\be\label{specurv}
\det (T_{N_c}(\mu)+w\cdot {\bf 1})=0
\ee
The coefficients in this equation can be taken as a complete set of
integrals of motion.

The manifest form of the curve in the $XXZ$ case can be derived
using explicit expression for the Lax matrix (\ref{l-gen}) and 
reads\footnote{In fact, the coefficient in front of $P_{N_c}$, instead of 2,
is equal to $2\cosh\prod_i^{N_c}e^{S_{0,i}}$. It is, however, the integral
of motion that can be put equal to any number. We always fix 
$\sum_i^{N_c}S_{0,i}$ to be zero.} 
\be\label{hrena}
w+{Q_{2N_c}\left(e^{2\xi}
\right)\over e^{2N_c\xi+2\sum_i\xi_i}w}=2e^{-N_c\xi-\sum_i\xi_i}P_{N_c}
\left(e^{2\xi}\right)=2e^{-N_c\xi-\sum_i\xi_i}\left(e^{2N_c\xi}+
\ldots+e^{2\sum\xi_i}\right)
\ee
Changing the variables $e^{2\xi}=\mu^2\equiv \lambda$, $w\to
e^{-N_c\xi-\sum_i\xi_i}w$, this curve can
be rewritten in the hyperelliptic form {\em in $\lambda$ variables}:
\be\label{sc2}
w+{Q_{2N_c}\left(\lambda\right)\over w}=2P_{N_c}\left(\lambda\right),\
\ \ Y^2=P_{N_c}^2(\lambda)-Q_{2N_c}(\lambda),\ \ \ Y\equiv
\2\left(w-{Q_{2N_c}\left(\lambda\right)\over w}\right)
\ee
while in terms of "true" spectral parameter
$\xi$ this curve looks considerably more tricky. However, one can work with
variable $\lambda$ taking into account that it lives on a cylinder
or sphere with two marked points.
In the equation (\ref{sc2}), the polynomial
\be\label{212}
Q_{2N_c}(\lambda)=\prod_i \left(\lambda^2
-2K_{i}e^{2\xi_i}\lambda+e^{4\xi_i}\right)\equiv
\prod_i\left(\lambda-e^{2m_i^{(+)}}\right)
\left(\lambda-e^{2m_i^{(-)}}\right)
\ee
defines the masses $m_i^{(\pm)}$ of the matter hypermultiplets. They turn out 
to be functions of the second Casimirs $K_{i}$ of the algebra (\ref{pois}) at 
$i$-th site and corresponding inhomogeneity to be compared with 
(\ref{mpm})
\be\label{mpm5}
m_i^{(\pm)}=\xi_i\pm{1\over 2}\log\left(K_{i}+\sqrt{K_{i}^2-1}\right)
\ee 
At the same time, the second
polynomial $P_{N_c}(\lambda)$ depends on the integrals of motion (mutually
Poisson-commuting with each other) playing the role of moduli of gauge theory
in the framework of the SW theory.
Modulo these subtleties, the curve (\ref{sc2}) is very similar to the curves
arising for $4d$ theories. However, the difference coming from different
generating 1-differentials $dS$ turns to be very crucial (see, for example, the
discussion of residue formula and perturbative prepotentials).

It is easy to compare the curves (\ref{sc2}) with those of \cite{theisen}, 
rewriting them in the form
\be\label{sinhsc}
w+{\prod_{\alpha}\sinh\left(\xi-m_{\alpha}\right)\over w}=2\prod_i
\sinh\left(\xi-a_i\right)
\ee
where we rescaled $w\to e^{N_c\xi+\sum_i\xi_i}w$,
denoted the roots of the polynomial $P_{N_c}(\lambda)$ (\ref{hrena})
$\lambda_i \equiv e^{2a_i}$ and made use of formula (\ref{mpm5}) and 
the manifest form of the leading and the constant terms 
in this polynomial. 
Comparing (\ref{sinhsc}) with (\ref{hrena}), one finds, in particular, that 
\be\label{strange}
\sum a_i=\sum\xi_i=\2\sum m_{\alpha}
\ee
The condition (\ref{strange}) is not, however, absolutely necessary in the
context of $XXZ$ chains and $5d$ theories (we return to this point later
in s.3.3).
It emerges only
in the standard $XXZ$ chain. In the context of SW theory one needs rather to 
consider the {\em twisted} $XXZ$ model \cite{kundu,ggm}
\footnote{We are grateful to S.Kharchev for the
discussion of the twisted spin chains.}. It is characterized 
by the Lax operator
\be\label{l-gentw}
L_i(\mu)\,=\,\left(
\begin{array}{cc}
\mu e^{S_{0,i}-\xi_i}-\alpha_i\mu^{-1}e^{-S_{0,i}+\xi_i} & 2S_{-,i}\\
2S_{+,i} & \mu e^{-S_{0,i}-\xi_i}-\beta_i\mu^{-1}e^{S_{0,i}+\xi_i}
\end{array}\right)
\ee
with generally non-unit constants $\alpha_i$, $\beta_i$. 
These constants provide arbitrary coefficient in front of 
product in the r.h.s. of (\ref{sinhsc}) (which can be also regulated by
the integral of motion $\sum_i^{N_c}S_{0,i}$) and, thus, break the condition 
(\ref{strange}). They are also important for careful treatment of dimensional 
transmutation (i.e. decreasing the number of massive hypermultiplets) 
\cite{ggm}.
The maximally degenerated case is related to the pure gauge theory when all the
masses become infinite. In this case, one gets
the relativistic Toda chain, first discussed in this context in 
\cite{nikita}.

There certainly exist all intermediate degenerations when all but $N_f$
masses are brought to infinity. This system, corresponding to the gauge theory
with $N_f<2N_c$ massive hypermultiplets, is described by the chain with some of
the sites degenerated and some of them not. The corresponding spectral curve
has the form
\be
w+{Q_{N_f}(\lambda)\over w}=2P_{N_c}(\lambda),\ \ \
Q_{N_f}(\lambda)=\prod_{\alpha}^{N_f}\left(\lambda-e^{2m_{\alpha}}\right),
\ \ \ P_{N_c}(\lambda)=\prod_i^{N_c}\left(\lambda-e^{2a_i}\right)
\ee
or
\be\label{scd5'}
w+{\prod_{\alpha}^{N_f}\sinh\left(\xi-m_{\alpha}\right)\over w}=
2e^{(N_c-N_f/2)\xi+\sum a_i-1/2\sum m_{\alpha}}\prod_i^{N_c}
\sinh\left(\xi-a_i\right)
\ee
(where we again rescaled $w\to e^{\2(N_f\xi+\sum m_{\alpha})}w$;
this curve also coincides with that in \cite{theisen}).
Below we mainly discuss the generic degenerated system with $N_f$ not
necessarily equal to $2N_c$.
As we already noted, the particular coefficient 
$e^{\sum a_i-1/2\sum m_{\alpha}}$ can be ,in principle, replaced by any other
number, say put equal to unity, which literally corresponds to \cite{theisen}. 
It does not influence the result and we will discuss the reason 
below.  

Note that the form (\ref{sinhsc})
of the spectral curve is perfectly designed for taking the 
$4d$ limit. Indeed, one can restore $R_5$-dependence in this formula 
multiplying each $\xi_i$ and mass parameter by $R_5$. In terms 
of algebra (\ref{pois}) it means that each generator has to be 
multiplied by $R_5$ and so does the spectral parameter. Then, one
immediately reproduces the results of sect.4.

\subsection{Relativistic integrable systems and $5d$ theories
\label{s:5d}}

The general scheme of the SW anzatz
presented above in the $4d$ case
can be almost literally transferred to the
``relativistic" $5d$ \N2 SUSY gauge models
with one compactified dimension. One can understand the
reason of this ``relativization'' in the following way. Considering
four plus one compact dimensional theory one should take into account the
contribution of all Kaluza-Klein modes to each 4-dimensional field.
Roughly speaking it leads to the 1-loop contributions to the effective
charge of the form
\be\label{relKK}
T_{ij} \sim \sum _{\rm masses}\log\hbox{ masses} \sim \sum _m\log
\left(a_{ij} +
{m\over R_5}\right) \sim \log\prod _m\left(R_5a_{ij} + m\right)
\sim\log\sinh R_5a_{ij}
\ee
i.e. coming from $4d$ to $5d$
one should make a substitution $a_{ij} \rightarrow 
\sinh R_5a_{ij}$, at least, in the
formulas for the perturbative prepotential. The similar change of variables
corresponds to relativization of integrable systems, which implies a
sort of "Lie group generalization" of the ``ordinary'' integrable systems
related to Lie algebras.
In terms of $2\times 2$ Lax representation, the ``relativization''
corresponds to replacing by trigonometric $r$-matrices and $L$-operators
the rational ones: in algebraic terms this corresponds to turning from
Yangians to quantum affine algebras. In particular, it means
that now it is natural to consider {\it both} parameters $\xi =
\2\log\lambda$
and $z=\log W$ as co-ordinates on cylinder, i.e. both {\em bare} spectral
curves become the cylinders.

Similar arguments imply that, instead of differential (\ref{dS}) 
$dS^{(4)}=\lambda d\log W$, one has now to consider the differential 
\cite{nikita,wdvv}
\be\label{dSRTC}
dS^{(5)} = \xi{dW\over W} =\2 \log\lambda {dW\over W}
\ee
so that, despite the similarities between the $5d$ and $4d$ spectral curves,
the periods of $dS^{(5)}$ are different from the periods of $dS^{(4)}$.

The difference with the $4d$ case results also in a slightly different
residue formula. Indeed, repeating the
derivation of the paper \cite{wdvv}, one can see that
\be
\frac{\partial^3 \CF}{\partial a_I\partial a_J\partial a_K} =2
\stackreb{{dz  = 0}\atop{\lambda=0}}{\res}
\frac{d\omega_Id\omega_Jd\omega_K}{dz  {d\lambda\over\lambda}}
\label{resfor5}
\ee
The crucial distinction is that the contour in (\ref{resfor5}) encircles
the ``marked point'' $\lambda = 0$ {\em together} with the zeroes of $dz$.
Using this formula, we derive in next section the perturbative
prepotential.

Thus, one can interpret the $5d$ theory as the $4d$ theory with infinite
number of (Kaluza-Klein) vector multiplets with masses $M_m={m\over R_5}$ and
infinite number of analogous (Kaluza-Klein) fundamental multiplets.
Then, the presented $5d$ picture naively corresponds to the infinitely
covered ``$4d$'' (see the spectral curve equation in $\xi$-variable  
(\ref{sinhsc}) living on a strip) 
also with infinitely many punctures. This, however, can be 
effectively encapsulated into hyperelliptic Riemann surface (\ref{sc2}) with
finite number of punctures, but, as a memory of infinitely many multiplets,
the spectral parameter $\lambda$ now lives on a cylinder (on a sphere
with two punctures). Meanwhile, the
differential $dS^{(5)}$ should be now  of the form (\ref{dSRTC})
which "remembers" its $4d$ origin.

Note that, in particular, the residue formula (\ref{resfor5}) can be written 
equally well in terms of variable $\xi$, instead 
of $\lambda$. Then, in the denominator of (\ref{resfor5}) there arises
the expression $dzd\xi$ and the residue is taken only at zeroes of $dz$
(note also that this change of variables eliminates the factor 2
in (\ref{resfor5})).
This form of the residue formula looks considerably more symmetric
and we use its analog for the $6d$ case in sect.4. 

\subsection{Perturbative limit of the SW anzatz}

In the perturbative limit, the hyperelliptic curve (\ref{sc2}) turns into the
rational curve
\be\label{pertcurve}
W=2{P_{N_c}(\lambda)\over\sqrt{Q_{N_c}(\lambda)}}\equiv
2{\prod_i^{N_c}\left(\lambda-\lambda_i\right)\over\sqrt{
\prod_{\alpha}^{N_f}\left(\lambda-\lambda_{\alpha}\right)}}
\ee
with $\lambda_i=e^{2a_i}$, $\sum_i^{N_c}a_i=0$, $\lambda_{\alpha}=
e^{2m_{\alpha}}$.
We consider here the case of arbitrary number $N_f\ge 2N_c$ of massive
hypermultiplets. First, let us construct the set of holomorphic differentials
$d\omega_I$.

The perturbative differential $dS$ is
\be\label{pertdS5}
dS =\2\log\lambda\  d\log\left(\frac{P(\lambda)}{\sqrt{Q(\lambda)}}\right)=
\2\log\lambda d\lambda
\left(\sum_{i=1}^{N_c}{1\over\lambda-\lambda_i}-\2\sum_{\alpha=1}
^{N_f}{1\over\lambda-\lambda_{\alpha}}\right)
\ee
or, equivalently,
\be\label{pertdS5xi}
dS=\xi d\xi\left(\sum_{i=1}^{N_c}\coth(\xi-a_i)-\2\sum_{\alpha=1}^{N_f}
\coth(\xi-m_{\alpha})+N_c-\2 N_f\right)
\ee
The variation of (\ref{pertdS5}) gives
\be\label{varpert}
2\delta dS = -\log\lambda d\left(\sum _{i=1}^{N_c}
{\delta\lambda _i\over\lambda - \lambda _i}  -\2\sum _{\alpha = 1}^{N_f}
{\delta\lambda _{\alpha}\over\lambda - \lambda _{\alpha}}\right) \cong
\nn \\
\cong {d\lambda\over\lambda} \sum _{i=1}^{N_c-1}
{\delta\lambda _i\over\lambda _i}\left( {\lambda _i\over\lambda - \lambda _i}-
{\lambda _{N_c}\over\lambda - \lambda _{N_c}}\right) -
{1\over 2}{d\lambda\over\lambda} \sum _{\alpha=1}^{N_f}
{\delta\lambda _{\alpha}\over\lambda _{\alpha}}\left(
{\lambda _{\alpha}\over\lambda - \lambda _{\alpha}}\right)
\ee
where it is implied that only $N_c -1$ parameters $\lambda _i$ are independent
so that ${\delta\lambda _{N_c}\over\lambda _{N_c}} = - \sum _{i=1}^{N_c-1}
{\delta\lambda _i\over\lambda _i}$. Now the set of the $N_c-1$ independent
canonical holomorphic differentials is
\be\label{difff5}
d\omega_i=
{d\lambda \over \lambda }\left( {\lambda _i\over\lambda - \lambda _i}-
{\lambda _{N_c}\over\lambda - \lambda _{N_c}}\right)
= {\lambda_{iN_c}d\lambda\over
(\lambda-\lambda_i)(\lambda-\lambda_{N_c})}\ \ \ 
i=1,...,N_c-1,
\ \ \ \lambda_{ij}\equiv \lambda_i-\lambda_j
\ee
and one can check that the ${\bf A}$-periods of $dS$
are $\log\lambda _i = 2a_i$, while the condition of vanishing 
the sum of all residues gives the relation (\ref{strange})
$\sum_i a_i=\2\sum m_{\alpha}$. In s.3.1, we discussed how this 
strange constraint should be removed off by using the twisted $XXZ$
spin chain. The analytical explanation of this is that one can easily
reproduce any lack of residues by putting some additional poles
at infinities in $\xi$ or at $\lambda=0,\infty$. It is analogous to
the procedure in $4d$. We shall see, however, that the same procedure
does not work in $d=6$.

The additional differentials corresponding to the derivatives of $dS$
(\ref{pertdS5}) with respect to masses have the form
\footnote{It can be instructive to 
describe in detail why (\ref{difff}) arises not only in $4d$, but also 
in $5d$ models. Strictly speaking, in the $5d$ case, 
one should consider 
differentials on annulus -- on sphere with 2 marked points. Therefore, 
instead of 
${d\lambda\over\lambda-\lambda_i}$, one rather needs to take 
\be\label{diffdr}
\sum_{m=-\infty}^{+\infty}{d\xi\over 
\xi-a_i+m}\sim\coth{(\xi-a_i)}d\xi
={\lambda+\lambda_i\over\lambda-\lambda_i}{d\lambda\over 2\lambda}
\ee
Taking now, instead of (\ref{difff}), the differentials (\ref{diffdr}) 
\be
d\omega_i={\lambda+\lambda_i\over\lambda-\lambda_i} 
{d\lambda\over 2\lambda} -{\lambda+\lambda_{N_c}\over\lambda-\lambda_{N_c}} 
{d\lambda\over 
2\lambda}= {\lambda_{iN_c}d\lambda\over 
(\lambda-\lambda_i)(\lambda-\lambda_{N_c})}
\ee
we obtain again the formula (\ref{difff}).
 On the other hand, the mass
differentials (\ref{massdiff5}) look different from (\ref{diffdr}). It turns
out, however, that the difference does not contribute into the results and,
therefore, one is free to choose any of these two possible mass
differentials. We will return to this point in next subsection.}
\be\label{massdiff5}
d\omega_{\alpha}=-{d\lambda\over 2\lambda}{\lambda_{\alpha}
\over \lambda -\lambda_{\alpha}}
\ee

\subsection{Perturbative prepotentials}

Now, using the manifest expressions for the differentials, we can
calculate the perturbative prepotentials using the residue formulas.
Let us first consider the case of pure $SU(N_c)$ $5d$ gauge theory,
when the prepotential is defined by the following residue formulas
\cite{wdvv}
\be
{\cal F}_{ijk}=2\stackreb{d\log P=0}{\res}
{d\omega_id\omega_jd\omega_k\over d\log
P(\lambda){d\lambda\over\lambda}}
\ee
since $\lambda=0$ does not contribute into the residues.
These residues can be easily calculated if one deforms the contour around
the poles of $d\omega_i$'s.
The final results have the following form
\be
{\cal F}_{iii}=\sum_{k\ne i}\coth a_{ik}+
6\coth a_{iN_c}+\sum_{k\ne N_c}\coth
 a_{kN_c},\\
{\cal F}_{iij}= -\coth a_{ij}+
4\coth a_{iN_c}+2\coth a_{jN_c}+\sum_{k\ne
i,j,N_c}\coth a_{kN_c}-N_c,\ \ \ i\ne j,\\
{\cal F}_{ijk}= 2\sum_{l\ne N_c}\coth a_{lN_c}-\sum_{l\ne i,j,k,N_c}\coth
 a_{lN_c}-N_c,\ \ \ i\ne j\ne k
\ee
From these expressions it follows
that the prepotential is given by formula
\be\label{FARTC}
{\cal F}={1\over 4}\sum_{i,j}\left({1
\over 3}a_{ij}^3
-{1\over 2}{\rm Li}_3\left(e^{-2a_{ij}}\right)\right)+
{N_c\over 2}\sum_{i>j>k}a_ia_ja_k=
\\={1\over 4}\sum_{i,j}\left({1
\over 3}a_{ij}^3
-{1\over 2}{\rm Li}_3\left(e^{-2a_{ij}}\right)\right)+
{N_c\over 6}\sum_{i}a_i^3
\ee
Now let us turn to the computation of the perturbative part of 
prepotential for
the $5d$ $SU(N_c)$ theory with $N_f$ massive fundamental hypermultiplets and
compare the results with the answers appeared in literature \cite{Sei}.
In order to include massive hypermultiplets, we
use the residue formula
\be\label{residueQ5}
{\cal F}_{IJK}=2\stackreb{{d\log {P\over\sqrt{Q}}=0}\atop{\lambda=0}}{\res}
{d\omega_Id\omega_Jd\omega_K\over d\log {P(\lambda)\over\sqrt{Q(\lambda)}}
{d\lambda\over \lambda}},
\ \ \ \ \
\left\{I,J,K,\dots\right\}=\left\{i,j,k,\ldots|\alpha ,\beta ,
\gamma ,\ldots\right\}
\ee
and obtain
\be\label{perth5}
\2{\cal F}_{iii}=\sum_{k\ne i}{\lambda_i\over \lambda_{ik}}+\
3{\lambda _i+\lambda_{N_c}\over
\lambda_{iN_c}}+ \sum_{k\ne N_c}{\lambda _{N_c}\over
\lambda_{kN_c}}+ \2\sum_{\alpha = 1}^{N_f}{\lambda_{iN_c}\lambda_{\alpha}\over
\lambda_{i\alpha}\lambda_{{N_c}{\alpha}}}
\\
\2{\cal F}_{iij}= 1+ 2{\lambda_{N_c}\over \lambda_{iN_c}}+
{\lambda _{N_c}\over \lambda_{jN_c}}+
\sum_{k\ne N_c}
{\lambda _{N_c}\over \lambda_{kN_c}}-
{\lambda _i\lambda_{jN_c}\over \lambda_{iN_c}\lambda_{ij}}+
\2\sum_{\alpha=1}^{N_f}{\lambda _{N_c}\over
\lambda_{N_c{\alpha}}},
\ \ \ i\ne j\\
\2{\cal F}_{ijk}=1+\sum_{l\ne N_c}{\lambda _{N_c}\over \lambda_{lN_c}} +
\sum _{l=i,j,k}{\lambda _{N_c}\over \lambda_{lN_c}} +
\2\sum_{\alpha =1}^{N_f}{\lambda _{N_c}\over\lambda_{N_c{\alpha}}},
\ \ \ i\ne j\ne k\\
{\cal F}_{ii{\alpha}}= {\lambda_{\alpha}\over\lambda_{i\alpha}}+
{\lambda_{\alpha}\over\lambda_{N_c\alpha}} \\
{\cal F}_{ij{\alpha}}={\lambda_{\alpha}\over\lambda_{N_c\alpha}}\ \ \ i\ne
j\\
{\cal F}_{i{\alpha}{\alpha}}={\lambda_{iN_c}\lambda_{\alpha}\over
\lambda_{i{\alpha}} \lambda_{N_c\alpha}}\\
{\cal F}_{\alpha\alpha\alpha} = 1 + \sum_{i=1}^{N_c}{\lambda_{\alpha}
\over\lambda_{i\alpha}} + \2\sum_{\beta = 1}^{N_f}{\lambda_{\alpha}
\over\lambda_{\alpha\beta}}\\
2{\cal F}_{\alpha\alpha\beta} =
-{\lambda_{\beta}\over\lambda_{\alpha\beta}} \\
{\cal F}_{\alpha\beta\gamma}={\cal F}_{i{\alpha}{\beta}}=0
\ee
i.e.
\be\label{perth1}
{\cal F}_{iii}=\sum_{k\ne i}\coth a_{ik}+
6\coth a_{iN_c}+\sum_{k\ne N_c}\coth
a_{kN_c}+\\+{1\over 2}\sum_{\alpha}
\left(\coth(a_{N_c}-m_{\alpha})-
\coth(a_{i}-m_{\alpha})\right)\\
{\cal F}_{iij}= -\coth a_{ij}+
4\coth a_{iN_c}+2\coth a_{jN_c}+\sum_{k\ne
i,j,N_c}\coth a_{kN_c}+{1\over 2}\sum_{\alpha}
\coth(a_{N_c}-m_{\alpha})-\\-N_c+{N_f\over 2},\ \ \ i\ne j\\
{\cal F}_{ijk}= 2\sum_{l\ne N_c}\coth a_{lN_c}-\sum_{l\ne i,j,k,N_c}\coth
a_{lN_c}+{1\over 2}\sum_{\alpha}
\coth(a_{N_c}-m_{\alpha})-\\-N_c+{N_f\over 2},\ \ \ i\ne j\ne k
\\
2{\cal F}_{ii{\alpha}}= \coth(a_i-m_{\alpha})+
\coth(a_{N_c}-m_{\alpha})-2
\\
2{\cal F}_{ij{\alpha}}=\coth(a_{N_c}-m_{\alpha})-1\ \ \ i\ne j
\\
2{\cal F}_{i{\alpha}{\alpha}}=(\coth(a_{N_c}-m_{\alpha})-
\coth(a_{i}-m_{\alpha}))
\\
2{\cal F}_{\alpha\alpha\alpha} =\sum_i\coth(a_i-m_{\alpha})+
\2 \sum_{\beta\ne\alpha}\coth(m_{\alpha}-m_{\beta})
-N_c+{N_f\over 2}+{3\over 2}
\\
4{\cal F}_{\alpha\alpha\beta} =\coth(m_{\beta}-m_{\alpha})
-1
\\
{\cal F}_{i{\alpha}{\beta}}=
{\cal F}_{\alpha\beta\gamma}= 0
\ee
The "strange" numbers -1, -2, ${3\over 2}$ that arise in the
derivatives of the prepotential w.r.t. the mass moduli are due to the choice
of the mass differentials in the form (\ref{massdiff5}). Certainly, these
numbers contribute into
at most quadratic part of the prepotential in gauge moduli. This does not
effect the "physical" implications but does break the WDVV equations 
\cite{wdvv}.
Meanwhile, with the second possible choice (see footnote 5)
\be\label{massdiff5'}
d\omega_{\alpha}=-{d\lambda\over 4\lambda}{\lambda+\lambda_{\alpha}\over
\lambda-\lambda_{\alpha}}=-{d\xi\over 2}\coth (\xi-m_\alpha)
\ee
these strange number are no longer present in the derivatives -- and this
turns out to be the only influence of the replace of the differentials.

The difference between the two forms of mass differentials is due to 
different possible choices of the factor in the r.h.s. of (\ref{scd5'}).
How it was already mentioned, this choice does not influence the result.
It does, however, the WDVV equations.

Thus, the $5d$ perturbative prepotential can be written in the following
plausible form
\be\label{prep5}
\CF={1\over 4}\sum_{i,j}f^{(5)}(a_{ij})-{1\over 4}\sum_{i,\alpha}
f^{(5)}(a_i-m_{\alpha})+{1\over 16}\sum_{\alpha,\beta}
f^{(5)}(m_{\alpha}-m_{\beta})+\\+{1\over 12}(2N_c-N_f)
\left(\sum_{i}a_i^3+{1\over 4}\sum_{\alpha}m_{\alpha}^3\right)
\ee
where, as in (\ref{f4d}) we use the functions $f^{(d)}(x)$ equal to
\be\label{fd}
f^{(5)}(x)={1\over 3}x^3
-{1\over 2}{\rm Li}_3\left(e^{-2x}\right)
\ee

\section{$XYZ$ chain and $6d$ theories}
\setcounter{equation}{0}
\subsection{General comments}

Now in the framework of general scheme developed in sect.2 and
applied to the $5d$ case in sect.3, we are going to turn to the $6d$ case,
or to the theory with {\em two} extra compactified dimensions,
of radii $R_5$ and $R_6$. The same argument as
we used for the transition from four to five dimensions, i.e. taking account
of the Kaluza-Klein modes, allows one to predict the perturbative form
of the charges in the $6d$ case as well. Namely, one should expect them
to have the form
\be\label{6dKK}
T_{ij} \sim \sum _{\rm masses}\log\hbox{ masses} \sim \sum _{m,n}\log
\left(a_{ij} +
{m\over R_5} +{n\over R_6}
\right) \sim 
\\
\sim\log\prod _{m,n}\left(R_5a_{ij} + m+n{R_5\over R_6}\right)
\sim\log\theta\left(R_5a_{ij}\left|i{R_5\over R_6}\right)\right.
\ee
i.e. coming from $4d$ ($5d$) to $6d$
one should replace the rational (trigonometric) expressions by the elliptic 
ones, at least, in the
formulas for the perturbative prepotential, the (imaginary part of) modular 
parameter being 
identified with the ratio of the compactification radii $R_5/R_6$.

A similar change of variables would give rise to `` elliptization'' of an 
integrable systems. In terms of $2\times 2$ Lax representation, the 
``elliptization''
corresponds to replacing by elliptic $r$-matrices and $L$-operators
the rational (\ref{laxmatr}) and trigonometric (\ref{l-gen}) one. In 
particular, it means
that now it is natural to consider parameters $\xi =
\2\log\lambda$
as a co-ordinate on torus, i.e. one of the bare spectral
curves becomes a torus, while the other one remains to be a cylinder.

In fact, we know the only system providing proper elliptization
of the $XXX$ spin chain, that is, the $XYZ$ Sklyanin chain \cite{Skl}
\footnote{In \cite{xyz} it has been already proposed that the $XYZ$ spin
chain should have something to do with the $N_f=2N_c$ SW theory. We will
see below, that this is indeed the case: the condition $N_f=2N_c$ which
can be easily broken in $4d$ and $5d$ situations is very strict here.
One may think of this theory as of blowing up all possible compactified
dimensions in the $N_f=2N_c$ theory with vanishing $\beta$-function, see
the discussion below.}. 
We propose
it here as a candidate for the integrable system behind the $6d$ theory.
Below we describe the $XYZ$ spin chain and demonstrate that it leads to the
perturbative prepotential which coincides with our 
naive expectations (\ref{6dKK}), giving some support to our conjecture. 

\subsection{$XYZ$ chain}

Now we briefly describe the $XYZ$ chain following \cite{xyz}. 
All necessary facts on the elliptic
functions can be found in Appendix. The Lax matrix
is now defined on elliptic curve
$E(\tau)$ and is explicitly given by (see \cite{FT} and references therein):
\be\label{39}
L^{\rm Skl}(\xi) = S^0{\bf 1} + i\frac{g}{\omega}\sum_{a=1}^3 W_a(\xi)S^a
\sigma_a
\ee
where
\be\label{sklyatheta}
W_a(\xi) = \sqrt{e_a - \wp\left({\xi}|\tau\right)} =
i\frac{\theta'_{\ast}(0)\theta_{a+1}\left({\xi}\right)}{\theta_{a+1}(0)
\theta_*(\xi)}
\ee
To keep the similarities
with \cite{xyz}, we redefine in this section the spectral
parameter $\xi\to i{\xi\over 2K}$, where $K\equiv\int_0^{{\pi\over
2}}{dt\over\sqrt{1-k^2\sin^2t}}={\pi\over 2}\theta_{00}^2(0)$,
$k^2\equiv{e_1-e_2\over e_1-e_3}$ so that $K\to{\pi\over 2}$ as $\tau\to
i\infty$. This factor results into additional multiplier $\pi$ in the
trigonometric functions in the limiting cases below.

The Lax operator (\ref{39})
satisfies the Poisson relation (\ref{quadr-r}) with the
numerical {\it elliptic} $r$-matrix $r(\xi)={i{g\over\omega}}\sum_{a=1}^3
W_a(\xi)\sigma_a\otimes\sigma_a$, which
implies that $S^0, S^a$ form the (classical)
Sklyanin algebra \cite{Skl}:
\be\label{sklyal}
\left\{S^a, S^0\right\} = 2i\left(\frac{g}{\omega}\right)^2
\left(e_b - e_c\right)S^bS^c
\nn \\
\left\{S^a, S^b\right\} = 2iS^0S^c
\ee
with the obvious notation: $abc$ is the triple $123$ or its cyclic
permutations.

The coupling constant  ${g\over \omega}$ can be eliminated by simultaneous
rescaling of the $S$-variables and the symplectic form:
\be
S^a =\frac{\omega}{g}\hat S^a \ \ \ \
S^0 = \hat S^0\ \ \ \
\{\ , \ \} \rightarrow -2\frac{g}{\omega}\{\ ,\ \}
\ee
Then
\be
L^{\rm Skl}(\xi) = \hat S^0 {\bf 1} +i\sum_{a=1}^3 W_a(\xi)\hat S^a\sigma_a
\ee
\be\label{sklyaln}
\left\{ \hat S^a,\ \hat S^0\right\} = -i
\left(e_c - e_b\right) \hat S^b\hat S^c \nn \\
\left\{ \hat S^a,\ \hat S^b\right\} = -i\hat S^0\hat S^c
\ee
One can distinguish three interesting
limits of the Sklyanin algebra: the rational, trigonometric and
double-scaling limits \cite{xyz}. We are interested in only
trigonometric limit here, since it describes the degeneration to the
$5d$ case. In this limit,
$\tau\rightarrow +i\infty$ and the Sklyanin algebra
(\ref{sklyaln}) transforms to
\be\label{trisklya}
\{ \hat S^3,\hat S^0\} = 0,\ \
\{\hat S^1,\hat S^0\} =i\hat S^2\hat S^3,\ \
\{\hat S^2,\hat S^0\} =-2i\hat S^3\hat S^1\\
\{\hat S^1,\hat S^2\} =-i\hat S^0\hat S^3,\ \
\{\hat S^1,\hat S^3\} = i\hat S^0\hat S^2,\ \
\{\hat S^2,\hat S^3\} = -i\hat S^0\hat S^1
\ee
The corresponding Lax matrix is
\be\label{laxxxz}
L_{XXZ} = \hat S^0{\bf 1}-{\f \sinh\pi\xi}\left(\hat S^1\sigma_1+
\hat S^2\sigma_2+\cosh\pi\xi\hat S^3\sigma_3\right)
\ee
and $r$-matrix
\be
r(\xi)={i\over\sinh\pi\xi}
\left(\sigma_1\otimes\sigma_1+\sigma_2\otimes\sigma_2+
\cosh\pi\xi\sigma_3\otimes\sigma_3\right)
\ee
Now one can note that the algebra (\ref{trisklya}) admits the identification
$\hat S_0=e^{S_0^{trig}}+e^{-S_0^{trig}}$,
$\hat S_3=e^{-S_0^{trig}}-e^{S_0^{trig}}$. With this identification and
up to normalization of the Lax operator (\ref{laxxxz}), we finally get the Lax
operator (\ref{l-gen}) of the $XXZ$ chain with the Poisson brackets
(\ref{pois}).

The determinant
$\det _{2\times 2} L^{\rm Skl}(\xi)$ is equal to
\be
\det _{2\times 2} L^{\rm Skl}(\xi) = \hat S_0^2 + \sum_{a=1}^3 e_a\hat S_a^2
- \wp(\xi)\sum_{a=1}^3\hat S_a^2
= K - M^2\wp(\xi) =  K - M^2 x
\ee
where
\be\label{casi}
K = \hat S_0^2 + \sum_{a=1}^3 e_a(\tau)\hat S_a^2 \ \ \
\ \ \ \
M^2 =  \sum_{a=1}^3 \hat S_a^2
\ee
are the Casimir operators of the Sklyanin algebra (i.e. Poisson
commuting
with all the generators $\hat S^0$, $\hat S^1$, $\hat S^2$,
$\hat S^3$). 

Now, in order to construct the spectral curve, note that 
the determinant of the monodromy matrix $T_{N_c}(\xi ) = 
\prod _{N_c\ge i\ge 1}^{\curvearrowleft}
L^{\rm Skl}(\xi - \xi_i)$ is
\be\label{54}
Q(\xi) = \det _{2\times 2} T_{N_c}(\xi) =
\prod_{i=1}^{N_c} \det _{2\times 2} L^{\rm Skl}(\xi - \xi_i) =
\prod_{i=1}^{N_c} \left( K_i - M^2_i\wp(\xi - \xi_i)\right) =
\\
= Q_0{\prod _{\alpha =1}^{N_f}\theta_{\ast}(\xi - m_\alpha)\over
\prod _{i=1}^{N_c}\theta _{\ast}(\xi - \xi_i)^2},\ \ \ 
Q_0=\prod_i^{N_c}A_i
\ee
where we used formulas (\ref{wptheta}), (\ref{e3}) and (\ref{thetaid}) 
from Appendix 
in order to rewrite $K_i-M_i^2\wp(\xi-\xi_i)$ as
$A_i{\theta_*\left(\xi-m_i^{(+)}\right)
\theta_*\left(\xi-m_i^{(-)}\right)\over \theta_*^2(\xi-\xi_i)}$. Here 
\be
A_i={M_i^2\left[\theta_*'(0)\right]^2\over \theta_*^2(\kappa_i)},\ \ \ 
m_i^{(\pm)}=\xi_i\pm \kappa_i
\ee 
and $\kappa_i$ is a solution to the equation
\be
{\theta_4^2(\kappa_i)\over\theta_*^2(\kappa_i)}=
{K_i\theta_4^2(0)\over M_i^2\left[\theta_*'(0)\right]^2}+
{\pi^2\theta_4^2(0)\left[\theta_2^4(0)+\theta_3^4(0)\right]\over 3
\left[\theta_*'(0)\right]^2}
\ee
Again
\be
\sum _{\alpha =1}^{N_f}m_\alpha = 2\sum _{i=1}^{N_c}\xi_i
\ee
As a particular example, let us consider 
the case of the {\it homogeneous} chain \cite{xyz} (all $\xi_i = 0$
in (\ref{54}))
$\Tr T_{N_c}(\xi)$ is a combination of the
polynomials:
\be\label{478}
P(\xi) = Pol^{(1)}_{\left[\frac{N_c}{2}\right]}(x) +
y Pol^{(2)}_{\left[\frac{N_c-3}{2}\right]}(x),
\ee
where $\left[\frac{N_c}{2}\right]$ is integral part
of ${N_c\over 2}$, and the coefficients of $Pol^{(1)}$ and $Pol^{(2)}$
are Hamiltonians of the $XYZ$ model.
The spectral equation (\ref{specurv}) for the $XYZ$ model
is:
\be\label{specXYZ}
w + \frac{Q(\xi)} w = 2P(\xi)
\ee
where for the homogeneous chain $P$ and $Q$
are polynomials in $x = \wp(\xi)$
and $y =   \frac{1}{2}\wp'(\xi)$.
Eq. (\ref{specXYZ}) describes the double covering of the elliptic
curve $E(\tau)$:
with generic point $\xi \in E(\tau)$ one associates the
two points of $\Sigma^{XYZ}$, labeled by two roots $w_\pm$
of equation (\ref{specXYZ}).
The ramification points correspond to
$w_+ =w_- = \pm\sqrt{Q}$, or
$Y = \frac{1}{2}\left(w - \frac{Q}w\right) =
\sqrt{P^2 - Q}= 0$. Note that, for the curve (\ref{specXYZ}),
$x = \infty$ is {\it not} a branch point,
therefore, the number of cuts on the both copies of $E(\tau)$ is $N_c$ and the
genus of the spectral curve is $N_c+1$.\footnote{Rewriting 
analytically $\Sigma^{XYZ}$ as a system of equations
\be
y^2 = \prod_{a=1}^3 (x - e_a), \ \ \ 
Y^2 = P^2 - Q
\ee
the set of holomorphic 1-differentials on $\Sigma^{XYZ}$ can be chosen as
\be\label{hdb}
v = \frac{dx}{y},\ \ \
V_\alpha = \frac{x^\alpha dx}{yY} \ \ \
\alpha = 0,\ldots,
\left[\frac{N_c}{2}\right], \ \ \ 
\tilde V_\beta = \frac{x^\beta dx}{Y} \ \ \
\beta = 0, \ldots,
\left[\frac{N_c-3}{2}\right]
\ee
with the total number of holomorphic 1-differentials
$1 + \left(\left[\frac{N_c}{2}\right] + 1\right) +
\left(\left[\frac{N_c-3}{2}\right] + 1\right) = N_c+1$
being equal to the genus of $\Sigma^{XYZ}$.}

In the case of a generic {\em inhomogeneous} chain the trace of monodromy
matrix looks more complicated, being expression of the form
\be\label{trsklya}
\Tr T_{N_c}(\xi ) = \Tr \prod _{N_c\ge i\ge 1}^{\curvearrowleft}
L^{\rm Skl}_i(\xi - \xi_i) =
\Tr\sum_{\{\alpha_i\}}\prod_{N_c\ge i\ge 1}^{\curvearrowleft}
\sigma^{\alpha_i}S^{\alpha_i}_i
W_{\alpha_i}(\xi-\xi_i) =
\\ = 
\sum_{\{\alpha_i\}}\Tr\prod_{N_c\ge i\ge 1}^{\curvearrowleft}
\sigma^{\alpha_i}S^{\alpha_i}_i
W_{\alpha_i}(\xi-\xi_i)
\\
\alpha_i = 0,a_i;\ \ \ \ \ \ a_i=1,2,3;\ \ \ \ \ W_0(\xi)=1 
\ee
It is clear from (\ref{sklyatheta}) that each
term in the sum (\ref{trsklya}) (and thus the whole sum) has simple poles in 
all inhomogeneities $\xi = \xi_i$. Thus the trace of the monodromy matrix
can be represented as
\be
P(\xi ) = {1\over 2}\Tr T_{N_c}(\xi ) = 
P_0{\prod_{i=1}^{N_c}\theta_{\ast}(\xi - a_i)\over
\prod_{i=1}^{N_c}\theta_{\ast}(\xi - \xi_i)}
\ee
\be\label{strange6}
\sum _i a_i = \sum _i\xi_i =\2\sum_{\alpha}m_{\alpha}
\ee
where $a_i$ should be identified with (shifted) gauge moduli 
-- see discussion in sect.3.1. 
The equation (\ref{specXYZ}) can be finally rewritten as
\be\label{specXYZ1}
w + Q_0{\prod _{\alpha =1}^{N_f}\theta_{\ast}(\xi - m_\alpha)\over
w\prod _{i=1}^{N_c}\theta _{\ast}(\xi - \xi_i)^2} = 
P_0{\prod_{i=1}^{N_c}\theta_{\ast}(\xi - a_i)\over
\prod_{i=1}^{N_c}\theta_{\ast}(\xi - \xi_i)}
\ee
We see that in $6d$ case it is necessary always to require $N_f=2N_c$ in
order to have a single-valued meromorphic functions on torus.
Again we also meet the constraint (\ref{strange6}). In the
$5d$ case, it was removed, along with the condition $N_f=2N_c$
twisting the spin chain. Since no twisted $XYZ$ chain is known
we are forced to preserve both constraints, their
origin will be discussed below
\footnote{Formula (\ref{strange6}) implies that, 
while the sum of all the would-be gauge moduli $a_i$ 
still remains a constant, it is no longer zero. Therefore,
one would rather associate with the gauge
moduli the quantities $a_i$ shifted by the constant ${1\over 2N_c}
\sum m_{\alpha}$. We do not know the proper field theory interpretation
of this condition.}.

The curve (\ref{specXYZ1})
has no longer that simple expression in terms of polynomials of
$x$ and $y$ similar to (\ref{478}).
Thus, in the inhomogeneous case it is not convenient to work in terms
of ``elliptic'' variables $x$ and $y$ (analogous to $\lambda$ in the $5d$ 
case) instead we choose to deal with the variable 
$\xi\in {\bf C}/\Gamma$. 

Given the spectral curve and integrable system one can immediately
write down the generating 1-differential $dS$.
It can be chosen in the following way
\be\label{dsxyz}
dS^{XYZ} \cong \xi {dw\over w}\cong - \log w d\xi
\ee
Return to our oversimplified example of the homogeneous chain.
Now, under the variation of moduli (which are all contained in $P$,
while $Q$ is moduli independent),
\be
\delta(dS^{XYZ}) \cong \frac{\delta w}{w}d\xi =
\frac{\delta P(\xi)}{\sqrt{P(\xi)^2-4Q(\xi)}} d\xi
= \frac{dx}{yY}\delta P
\ee
and, according to (\ref{hdb}), the r.h.s. is a {\it holomorphic}
1-differential on the spectral curve (\ref{specXYZ}).

The residue formula in the $6d$ case
is of the form completely analogous to the
$4d$ case (and to the $5d$ case, when working in terms of the variable
$\xi$)
\be
\frac{\partial^3 \CF}{\partial a_I\partial a_J\partial a_K} =
\stackreb{dz  = 0}{\res}
\frac{d\omega_Id\omega_Jd\omega_K}{dz d\xi}
\label{resfor6}
\ee
and we use it in the next subsections to derive the perturbative prepotential.

\subsection{Perturbative limit of the XYZ spin chain}

A generic $XYZ$ curve (\ref{specXYZ1}) in the perturbative limit
turns into
\be\label{pertXYZ}
W = h_0{\prod_{i=1}^{N_c}\theta_{\ast}(\xi - a_i)\over
\sqrt{\prod _{\alpha =1}^{N_f}\theta_{\ast}(\xi - m_\alpha)}}
\ee
where $h_0 = {2P_0\over\sqrt{Q_0}}$. Then, 
the rescaled generating differential (\ref{dsxyz}) is 
\be\label{pertdS6}
dS= \left(\sum\log\theta_{\ast}(\xi-a_i) - \2\sum\log\theta_{\ast}(\xi-m_a) 
\right) d\xi + \log h_0d\xi\cong\\
\cong
\xi d\xi\left(\sum_{i=1}^{N_c} \zeta (\xi - a_i) - 
\2\sum_{\alpha=1}^{N_f}\zeta (\xi - m_\alpha)\right)
\ee
where the linear piece disappears due to $\sum a_i=\2\sum m_\alpha$. This
expression should be compared with (\ref{pertdS5xi}) in the $5d$ case.

Note that the sum of residues vanishes only provided the constraint
(\ref{strange6}) is fulfilled. It can not be removed by adding an
additional pole at infinity, since the variable $\xi$, in contrast to
the $4d$ and $5d$ cases, lives on the compact surface with out
boundaries (torus) and, therefore, is always finite. 

Now one can take the
total variation of $dS$ (we neglect the trivial variation $\delta h_0$
giving rise to the holomorphic differential $d\xi $ on the bare torus)
\be
\delta dS^{XYZ} \cong \sum_{i=1}^{N_c} \bar\zeta (\xi - a_i)\delta a_id\xi - 
\2\sum_{\alpha=1}^{N_f}\bar\zeta (\xi - m_\alpha)\delta m_\alpha d\xi 
\ee
where $\bar\zeta(\xi|\tau)$ is defined in Appendix (see (\ref{zeta'})).
Therefore, the differentials related to gauge moduli are
\be\label{diff6}
d\omega_i=\left(\bar\zeta(\xi-a_i)-\bar\zeta(\xi-a_{N_c})\right)d\xi
\ee
while those related to masses are
\be\label{massdiff6}
d\omega_\alpha=-\2\left(\bar\zeta(\xi-m_\alpha)
-\bar\zeta(\xi-a_{N_c})\right)d\xi
\ee
The second term in this expression is due to the condition 
(\ref{strange6}). Thus, one can see that, in the $6d$ case, mass moduli
are practically identical to the gauge ones. 

The residue formula acquires the form
\be\label{resklya}
\CF_{IJK} = \res _{d\log {P\over\sqrt{Q}}=0} {d\omega_Id\omega_Jd\omega_K
\over d\xi d\log {P\over\sqrt{Q}}}
\ee

\subsection{Perturbative prepotentials}

Having the explicit formulas for the differentials (\ref{diff6}) and 
(\ref{massdiff6}), one can use the residue formula (\ref{resklya})
to calculate the perturbative prepotential in the $6d$ case. The result is
\be
{\cal F}_{iii}=\sum_{k\ne i}\bar\zeta( a_{ik})+
6\bar\zeta (a_{iN_c})+\sum_{k\ne N_c}\bar\zeta
(a_{kN_c})+\\+{1\over 2}\sum_{\alpha}
\left(\bar\zeta(a_{N_c}-m_{\alpha})-
\bar\zeta(a_{i}-m_{\alpha})\right)\\
{\cal F}_{iij}= -\bar\zeta (a_{ij})+
4\bar\zeta (a_{iN_c})+2\bar\zeta (a_{jN_c})+\sum_{k\ne
i,j,N_c}\bar\zeta (a_{kN_c})+{1\over 2}\sum_{\alpha}
\bar\zeta(a_{N_c}-m_{\alpha}),\ \ \ i\ne j\\
{\cal F}_{ijk}= 2\sum_{l\ne N_c}\bar\zeta (a_{lN_c})-
\sum_{l\ne i,j,k,N_c}\bar\zeta
(a_{lN_c})+{1\over 2}\sum_{\alpha}
\bar\zeta(a_{N_c}-m_{\alpha}),\ \ \ i\ne j\ne k\\
2{\cal F}_{ii{\alpha}}= \bar\zeta(a_i-m_{\alpha})+
\bar\zeta(a_{N_c}-m_{\alpha})+\bz(a_{N_ci})+\SUM\\
2{\cal F}_{ij{\alpha}}=\bar\zeta(a_{N_c}-m_{\alpha})
+\bz (a_{N_ci})+\bz (a_{N_cj})+\SUM\ \ \ i\ne j\\
2{\cal F}_{i{\alpha}{\alpha}}=\bar\zeta(a_{i}-m_{\alpha})
+\2\bz (a_{iN_c})-\2\left(\SUM\right)\\
2{\cal F}_{\alpha\alpha\alpha} =\sum_i\bar\zeta(a_i-m_{\alpha})+
\2 \sum_{\beta\ne\alpha}\bar\zeta(m_{\alpha}-m_{\beta})
+{1\over 4}\left(\SUM\right)+\\
+{3\over 4}\bz (m_\alpha -a_{N_c})\\
4{\cal F}_{\alpha\alpha\beta} =\bar\zeta(m_{\beta}-m_{\alpha})
+\2\bz (m_\alpha -a_{N_c})+\2\left(\SUM\right)\\
4{\cal F}_{i{\alpha}{\beta}}=\bz (m_\alpha -a_{N_c})+\bz (m_\beta -a_{N_c})
+\bz (a_{iN_c})-\SUM\\
8{\cal F}_{\alpha\beta\gamma}= -\bz (m_\alpha -a_{N_c})-\bz (m_\beta -a_{N_c})
-\bz (a_{iN_c})+\SUM
\ee
These formulas imply that the perturbative prepotential is given by
\be
\CF={1\over 4}\sum_{i,j}f^{(6)}(a_{ij})-{1\over 4}\sum_{i,\alpha}
f^{(6)}(a_i-m_{\alpha})+{1\over 16}\sum_{\alpha,\beta}
f^{(6)}(m_{\alpha}-m_{\beta})
\ee
where $\sum a_i=\2\sum m_\alpha$ and
the function $f^{(6)}(x)$ defined so that
\be
{f^{(6)}}''(x)=\log\theta_{\ast}(x)
\ee
can be expressed through 
the elliptic tri-logarithm ${\rm Li}_{3,q}(x)$ 
\cite{mamont}:
\be\label{f6}
f^{(6)}(a)=\sum_{m,n}f^{(4)}\left(a+\frac{n}{R_{5}}+\frac{m}{R_{6}}\right)=
\sum_n f^{(5)}\left(a+n{R_{5}\over R_6}\right)=
\\
= \left({1
\over 3}a^3-{a^2\over 4}+{1\over 120}
-{1\over 2}{\rm Li}_{3,q}\left(e^{-2a}\right)\right)
\ee

\section{Discussion}
\setcounter{equation}{0}
Now we are going to discuss the results. 
We have reproduced the spectral curves for the SUSY $SU(N_c)$ gauge theories 
with $N_f$ massive hypermultiplets in $d=4,5$ and 6 dimensions. Using these 
curves and the corresponding differentials, we have calculated the 
perturbative parts of the prepotentials in all these cases and the results
have the same form.

\paragraph{Results.}
The spectral curve in all cases can be written in the form
\be\label{scg}
w+{Q^{(d)}(\xi)\over w}=2P^{(d)}(\xi)
\ee
or
\be\label{scg'}
W+{1\over W}={2P^{(d)}(\xi)\over\sqrt{Q^{(d)}(\xi)}},
\ \ \ \ \ \ W = {w\over\sqrt{Q^{(d)}(\xi)}}
\ee
The generating differential $dS$ is always of the form
\be
dS=\xi d\log W
\ee
The perturbative part of the prepotential is always of the form
\be
\CF={1\over 4}\sum_{i,j}f^{(d)}(a_{ij})-{1\over 4}\sum_{i,\alpha}
f^{(d)}(a_i-m_{\alpha})+{1\over 16}\sum_{\alpha,\beta}
f^{(d)}(m_{\alpha}-m_{\beta})
\ee
The concrete forms of the functions introduced here are:
\be
Q^{(4)}(\xi)\sim\prod_\alpha^{N_f}(\xi-m_\alpha),\ \ \ 
Q^{(5)}(\xi)\sim\prod_\alpha^{N_f}\sinh(\xi-m_\alpha),\ \ \ 
Q^{(6)}(\xi)\sim\prod_\alpha^{N_f}{\theta_*(\xi-m_\alpha)\over\theta_*^2
(\xi-\xi_i)}
\ee
\be
P^{(4)}\sim\prod_i^{N_c}(\xi-a_i),\ \ \ 
P^{(5)}\sim\prod_i^{N_c}\sinh(\xi-a_i),\ \ \ 
P^{(6)}\sim\prod_i^{N_c}{\theta_*(\xi-a_i)\over\theta_*(\xi-\xi_i)} 
\ee
(in $P^{(5)}(\xi)$, there is also some exponential of $\xi$ unless $N_f=2N_c$,
see (\ref{scd5'}))
\be
f^{(4)}(x)=x^2\log x,\ \ \
f^{(5)}(x)=\sum_{n}f^{(4)}\left(x+{n\over R_5}\right)=
{1\over 3}\left|x^3
\right|-{1\over 2}{\rm Li}_3\left(e^{-2|x|}\right),\\
f^{(6)}(x)=\sum_{m,n}f^{(4)}\left(x+\frac{n}{R_{5}}+\frac{m}{R_{6}}\right)=
\sum_n f^{(5)}\left(x+n{R_{5}\over R_6}\right)=
\\
= \left({1
\over 3}\left|x^3\right|
-{1\over 2}{\rm Li}_{3,q}\left(e^{-2|x|}\right)
+ {\rm quadratic\ \ terms}\right)
\ee
so that
\be
{f^{(4)}}''=\log x,\ \ \ {f^{(5)}}''(x)=\log\sinh x,\ \ \ 
{f^{(6)}}''(x)=\log\theta_*(x)
\ee
Note that, in the $6d$ case, $N_f$ is always equal to $2N_c$. The variables
$\xi_i$ above are inhomogeneities in integrable system, and, in $d=5,6$,
there is a restriction $\sum a_i=\sum\xi_i=\2\sum m_\alpha$ 
which implies that the gauge moduli would be
rather associated with $a_i$ shifted by the constant 
${1\over 2N_c}\sum m_\alpha$.

\paragraph{On perturbative limit.}
Now let us make some comments concerning the perturbative limit in all 
these cases. In fact, the procedure is quite clear in the $4d$ and $5d$
cases: this limit corresponds to big values of gauge moduli (and
masses) compare to $\Lambda_{QCD}$ and can be taken as 
${\Lambda_{QCD}\over \mu}\to 0$, which is exactly the perturbative limit in 
asymptotically free theories.
Looking at (\ref{scg'}), one can see that
this procedure corresponds to $\xi$ going to
infinity (as $\Lambda_{QCD}^{-1}$) and $W$ going to infinity 
or to zero depending on the choice of a particular list
of hyperelliptic spectral curve
\footnote{In $4d$ $W\to \infty$ as $\Lambda_{QCD}^{\2 N_f-N_c}$. 
However, one can renormalize the gauge moduli,
masses, $W$ and $\xi$ so that they remain finite as $\Lambda_{QCD}\to 0$.
It results into the coefficient $\Lambda_{QCD}^{2N_c-N_f}$ 
in front of the term ${1\over W}$ in (\ref{scg'}).
The analogous procedure can be also performed for the other sheet, when 
$W\to 0$.
In $5d$ the dependence on $\Lambda_{QCD}$ can not be reduced to the
only coefficient, this is an indication of non-renormalizability of $5d$ 
theory.}. 
From the point of view of the effective charges $T_{ij}$ it
separates logarithmic terms from the power (instantonic) contributions.

This reasoning certainly fails when $d=6$. Indeed, since 
$\xi\in{\bf C}/\Gamma$ it can not be taken to infinity. In terms of 
prepotential it means that one can not distinguish between $\log\theta$
terms and the other since $\theta$-function is restricted.
However, there is a way out of this problem, although a little bit
artificial. Namely, instead of moving $\xi$ one may just turn
to zero the coefficient in front of the product in $Q^{(6)}(\xi)$. 
The perturbative contribution to the $6d$ prepotential 
calculated in the paper correspond exactly to this limit.

As it is well-known, $5d$ and $6d$ theories are 
non-renormalizable. Therefore, the status of the perturbative 
contributions has to be understood better (see also \cite{Sei,d6,nikita,nl}). 
We are not going to discuss this question in detail here, instead let us 
point out that the problems of
renormalizability are related only with maximally {\em quadratic} part
of the prepotential getting stronger than logarithmic divergencies in $5d$ and
$6d$ perturbative theory. Our approach does not control these terms and
the properties of (compactified) $5d$ and $6d$ theories we discuss do
not depend on quadratic contributions.

The first property is the periodicity of the $5d$ prepotential under 
$a_i\to a_i+2i\pi n{R_5}^{-1}$, with integer $n$, which is evidently correct 
up to quadratic terms (being changed due to the shifts in cubic terms). 
This shift is induced by changing the sheet of the logarithm in 
$dS^{(5)}$ (\ref{dSRTC}) (either $\log\lambda$, or
$\log w$, depending on the chosen form of $dS^{(5)}$) as one can easily
obtain from (\ref{defprep}). It 
explains, in particular, why the ambiguity in definition of $dS^{(5)}$ 
(\ref{dSRTC}) does not influence the result.
Similarly, in $6d$ case one needs double-periodicity to guarantee
the unambiguous prepotential despite the multi-valuedness of 
generating differential $dS^{(6)}$ (\ref{dsxyz}). 
This is indeed the case: $f^{(6)}$ in (\ref{f6}) is a double-periodic function.

The second property of the prepotentials is typically to satisfy the
WDVV equations \cite{wdvv1,wdvv}. This is strongly
related to integrable structure of the theory and
should have a topological origin. The WDVV equations
are equations for the {\em third} derivatives and do not depend on
quadratic terms. On the other hand, these 
equations requires to consider masses and gauge moduli on equal footing. 
Therefore, the WDVV equations are sensible to the
constant corrections to the third derivatives of the prepotential, even
if some of these derivatives are taken w.r.t. masses.

This issue has a lot of to do with the proper definition of mass
differentials (see sect.3), we are going to return to this problem
elsewhere.

\paragraph{$5d$ and $6d$ theories -- cubic terms, $N_f=2N_c$ etc.}
Next point we are going to discuss is the cubic (in moduli) terms 
appearing only in higher-dimensional theories.
First, take the $5d$ case, where these terms corresponds to
the Chern-Simons term in the field theory Lagrangian 
$\Tr (A\wedge F\wedge F)$. In our computation they appear due to the
constant piece $(N_c-\2 N_f)$ in the brackets in (\ref{pertdS5xi}).

In fact, let us
choose the special values of the second Casimir functions at $m$ sites so that
the polynomial $Q(\lambda)$ acquires the form 
$\lambda^m\bar Q_{N_f}(\lambda)$, i.e. $m$ of $2N_c$
factors in (\ref{212}) turn into $\lambda^m$. In this case one gets 
the constant piece $N_c-m-N_f/2$ in $dS^{(5)}$,
with an integer $m\le 2N_c-N_f$. Therefore, the coefficient in front of 
cubic term in (\ref{prep5}) can be made equal to ${1\over 6}(N_c-m-N_f/2)$.

Restoring the dependence on $R_5$ in (\ref{prep5}) ($a_i\to
a_iR_5$ and $m_{\alpha}\to m_{\alpha}R_5$) one can study the different
limits of the system.
The simplest limit corresponds to $4d$ case and is given by
$R_5\to 0$. In this limit, $f^{(5)}(x)
\stackreb{x\sim 0}{\to}f^{(4)}(x)$, the cubic terms vanish 
and we reproduce the perturbative $4d$ prepotential (\ref{f4d}).
At the level of
integrable system it is enough to replace $S_i\to R_5S_i$,
$\mu\to e^{R_5\mu}$ in the Lax operator (\ref{l-gen}) ((\ref{l-gentw})) in
order to reproduce the Lax operator of the $XXX$ (higher $sl(p)$) spin chain, 
see \cite{ggm} for details).

Another interesting limit is the limit of flat $5d$ space-time, i.e.
$R_5\to\infty$. In this limit, only cubic terms survive in the prepotential
(\ref{prep5}) (one should carefully fix the branch of $f^{(6)}(x)$ which 
leading to appearing of the absolute value in
(\ref{seith})):
\be\label{seith}
\CF={1\over 12}\sum_{i,j}\left|a_{ij}\right|^3-{1\over 12}\sum_{i,\alpha}
\left|a_i+m_{\alpha}\right|^3+{1\over 48}\sum_{\alpha,\beta}
\left|m_{\alpha}-m_{\beta}\right|+\\
+{1\over 12}(2N_c-N_f)
\left(\sum_{i}a_i^3+{1\over 4}\sum_{\alpha}m_{\alpha}^3\right)
\ee
In fact, there are two different sources
of cubic terms \cite{Sei}. 
The first one is from the function $f^{(5)}(x)$.
Since this function can be obtained as the sum of the $4d$
perturbative contributions over the Kaluza-Klein modes,
these cubic terms have perturbative origin and come from the 1-loop (due to
the well-known effect of generation of the CS terms in odd-dimensional gauge
theories).

The second source of the cubic terms is due to the bare CS Lagrangian. As it
was shown in \cite{Sei}, one can consider these terms with some
coefficient $c_{cl}$:
\be
{c_{cl}\over 6}\sum_i a^3_i
\ee
restricted only to satisfy the quantization condition $c_{cl}+{N_f\over
2}\in \Z$ and the inequality $\left|c_{cl}\right|\ge N_c-{N_f\over 2}$.
In our formulas, we easily reproduce these conditions so that
(\ref{seith}) coincides with \cite{Sei}, provided
the prepotential is
defined in a fixed Weyl chamber. We leave, however, the wall crossing jumps
out of the discussion.
Note that the curve (\ref{sc2}) with this arbitrary $m$
coincides with that proposed in \cite{theisen}.

Note also that it is clear why the WDVV equations are not necessarily 
satisfied when $m\ne 0$. Since it corresponds to degeneration of an 
integrable system, one fixes some moduli (masses) -- the parameters 
to vary in the WDVV equations (see \cite{wdvv} for details). 

In the $6d$ case the calculations are quite similar, and one would expect 
the similar cubic terms correspond to $\Tr (F\wedge F\wedge F)$. However, 
despite all the difference
in calculations is due to slightly different mass differentials
(\ref{massdiff5'}) and (\ref{massdiff6})
the bare cubic terms in $6d$ are absent. This occurs due to
$N_f=2N_c$ (since the coefficient in front of cubic terms is
proportional to $N_c-\2 N_f$) and, therefore, 
cancelation of the constant term in the brackets of (\ref{pertdS6}) 
as compared with (\ref{pertdS5xi}).
This is one of the problems with $6d$ theories: one can not
deform (twist) the theory so that the constraints $N_f=2N_c$ and
(\ref{strange6}) look unavoidable. We do not understand clearly
their meaning at the level of field theory.
As for the ``quantum generated" terms, i.e. those coming from the function
$f^{(6)}$, they are certainly presented, with the coefficient equal to that
in the $5d$ case.

Finally, let us point out that the Lax matrices (\ref{laxmatr}), (\ref{l-gen})
and (\ref{39}) (and analogously though more involved for the transfer matrix
(\ref{trsklya})) can be possibly determined from a linear differential
equations $\bar D L_i(\xi ) \sim \delta^{(2)}(\xi-\xi_i)$ 
({\it a l\'a} Hitchin). These equations could be interpreted as arising in the 
framework of D-brane interpretation of SUSY gauge theories \cite{W,MMaM}. 
It might indicate that the standard Hitchin-like systems and the 
spin-chain like models, in spite of distinctions in the formulation
of the Poisson structures etc \cite{FT}, can be treated within some unified 
approach in the framework of higher-dimensional
SUSY gauge theories and corresponding brane configurations of string theory.
We are going to return to this problem elsewhere.

\section*{Appendix}
\def\theequation{A.\arabic{equation}}
\setcounter{equation}{0}

The definition of elliptic functions we use throughout the paper slightly
differs from standard definitions that can be found, for example, in 
\cite{WW}. In this Appendix we define all the
functions we need and write down manifestly some useful identities
necessary for sect.4.

We define the Weierstrass $\wp$-function to be the sum
\be\label{A1}
\wp(\xi|\tau)=\sum_{m,n=-\infty}^{+\infty}{1\over (\xi+m+n\tau)^2}-
{\sum_{m,n=-\infty}^{+\infty}}'
{1\over (m+n\tau)^2}={1\over\xi^2}+{\cal O}(\xi^2)
\ee
where $\sum '$ means omitted term with $m=n=0$. Thus defined $\wp$-function
is double periodic, with periods 1 and $\tau={\omega\over\omega '}$ and
differs from the standard definition by the factor $(2\omega)^{-2}$ and
rescaling $\xi\to 2\omega\xi$. According to this definition, the values of
$\wp(\xi|\tau)$ in the half-periods, $e_a=e_a(\tau)$, $a=1,2,3$, also differ
by the factor $(2\omega)^{-2}$ from the standard definition.

The complex torus $E(\tau)$ can be defined as $\C/\Z\oplus\tau\Z$ with
a flat co-ordinate $\xi$ defined modulo $(1,\tau)$. Alternatively,
any torus (with a marked point) can be described as the elliptic curve
\be\label{A2}
y^2=(x-e_1)(x-e_2)(x-e_3),\ \ \ x=\wp(\xi),\ \ \ y=\2\wp '(\xi)
\ee
and the canonical holomorphic 1-differential is
\be\label{A3}
d\xi=2{dx\over y}
\ee
Other elliptic functions used in the paper are $\zeta$-function
\be\label{zeta}
\zeta(\xi|\tau)={1\over\xi}+{\sum_{m,n=-\infty}^{+\infty}}'\left(
{1\over \xi -m-n\tau}+{1\over m+n\tau}+{\xi\over (m+n\tau)^2}\right)=
{1\over\xi}+{\cal O}(\xi^3)
\ee
and $\theta$-functions
\be\label{theta}
\theta_1(\xi|\tau)=\theta_{11}(\xi|\tau)=\theta_{*}(\xi|\tau)=
i\sum_{n=-\infty}^{+\infty}(-1)^ne^{i(n-\2)^2\pi\tau}e^{i\pi(2n-1)\xi}\\
\theta_2(\xi|\tau)=\theta_{01}(\xi|\tau)=
\sum_{n=-\infty}^{+\infty}e^{i(n-\2)^2\pi\tau}e^{i\pi(2n-1)\xi}\\
\theta_3(\xi|\tau)=\theta_{00}(\xi|\tau)=
\sum_{n=-\infty}^{+\infty}e^{in^2\pi\tau}e^{i\pi 2n\xi}\\
\theta_4(\xi|\tau)=\theta_{10}(\xi|\tau)=
\sum_{n=-\infty}^{+\infty}(-1)^ne^{in^2\pi\tau}e^{i\pi 2n\xi}
\ee
There exist the following relations connecting different elliptic
functions and constants with
$\theta$-functions and $\theta$-constants $\theta(0)\equiv\theta(0|\tau)$:
\be\label{zeta'}
\zeta(\xi)=\eta(\tau)\xi+\left[\log\theta_*(\xi)\right]'
\ee
where
\be\label{eta}
\eta(\tau)=-{1\over 6}{\theta_*'''(0)\over\theta_*'(0)}
\ee
\be\label{wptheta}
\wp(\xi)=-\zeta '(\xi)=e_3+\left[{\theta_*'(0)\over\theta_4(0)}
{\theta_4(\xi)\over\theta_*(\xi)}\right]^2
\ee
with
\be\label{e3}
e_3(\tau)=-{\pi^2\over 3}\left[\theta_2^4(0)+\theta_3^4(0)\right]
\ee
To simplify some formulas, we also introduce the logarithmic derivative
of the $\theta$-function
\be\label{barzeta}
\bar\zeta(\xi|\tau)\equiv\zeta(\xi|\tau)-
\eta(\tau)\xi=\left[\log\theta_*(\xi)\right]'
\ee
In sect.4, we also use the identity for $\theta$-functions \cite{WW}
\be\label{thetaid}
\theta_*(x+y)\theta_*(x-y)\theta_4^2(0)=\theta_*^2(x)\theta_4^2(y)-
\theta_4^2(x)\theta_*^2(y)
\ee

\section*{Acknowledgements}
We are indebted to A.Gorsky, S.Kharchev, A.Zabrodin and especially
to A.Morozov for illuminating discussions. A.Mar. is also grateful to J.Ambjorn
and other members of the Theoretical High-Energy group of the Niels Bohr 
Institute where this work was completed, and A.Mir. acknowledges the 
University Roma I for warm hospitality. The work of A.Mar. was partially 
supported by 
RFBR grant 96-01-01106, INTAS grant 96-518 and the Niels Bohr Institute, 
the work of A.Mir. -- by the grant RFBR-96-02-16347(a), the program for 
support of the scientific schools 96-15-96798 and INTAS-96-482.


\begin{thebibliography}{12}

\bibitem{SW}
N.Seiberg and E.Witten, ``Electric-Magnetic Duality,
Monopole Condensation, And Confinement in $N=2$ Supersymmetric
Yang-Mills Theory '', Nucl.Phys. {\bf B426} (1994) 19-52; Erratum-ibid.
{\bf B430} (1994) 485-486; hep-th/9407087\\
``Monopoles, Duality and Chiral Symmetry Breaking in
N=2 Supersymmetric QCD'',
Nucl.Phys. {\bf B431} (1994) 484-550; hep-th/9408099

\bibitem{GKMMM}
A.Gorsky, I.Krichever, A.Marshakov, A.Mironov and A.Morozov,
``Integrability and
Seiberg-Witten Exact Solution'',
Phys.Lett. {\bf B355} (1995) 466-474; hep-th/9505035

\bibitem{SWIS}
E.Martinec and N.Warner, ``Integrable systems and supersymmetric gauge theory",
Nucl.Phys. {\bf B459} (1996) 97-112; hep-th/9509161;\\
T.Nakatsu and K.Takasaki, ``Whitham hierarchy and N=2 supersymmetric 
Yang-Mills theory", Mod.Phys.Lett. {\bf A11} (1996) 157-168; 
hep-th/9509162;\\
R.Donagi and E.Witten, ``Supersymmetric Yang-Mills Systems And Integrable
Systems", Nucl.Phys. {\bf B460} (1996) 299; hep-th/9510101;\\
E.Martinec, ``Integrable structures in supersymmetric gauge and string theory",
Phys.Lett. {\bf B367} (1996) 91-96; hep-th/9510204;\\
A.Gorsky and A.Marshakov, ``Towards effective topological gauge theories
on spectral curves", Phys.Lett. {\bf B375} (1996) 127, hep-th/9510224;\\
E.Martinec and N.Warner, ``Integrability in N=2 gauge theory: a proof",
hep-th/9511052;\\
H.Itoyama and A.Morozov, ``Integrability and Seiberg-Witten Theory: Curves
and Periods", Nucl.Phys. {\bf B477} (1996) 855-877; hep-th/9511126;
``Prepotential and the Seiberg-Witten Theory", Nucl.Phys. {\bf B491} (1997)
529-573; hep-th/9512161;\\
C.Ahn and S.Nam, ``Integrable structure in supersymmetric gauge theories
with massive hypermultiplets", Phys.Lett. {\bf B387} (1996) 304-309;
hep-th/9603028;\\
I.Krichever and D.Phong, ``On the integrable geometry of soliton equations
and N=2 supersymmetric gauge theories", J.Diff.Geom. {\bf 45} (1997)
349-389; hep-th/9604199;
``Symplectic forms in the theory of solitons", hep-th/9708170;\\
T.Nakatsu and K.Takasaki, ``Isomonodromic deformations and supersymmetric
gauge theories", Int.J.Mod.Phys. {\bf A11} (1996) 5505-5518; 
hep-th/9603069;\\
K.Takasaki, ``Spectral Curves and Whitham Equations in Isomonodromic 
Problems of Schlesinger Type",  solv-int/9704004;
``Dual Isomonodromic Problems and Whitham Equations", 
solv-int/9705016;\\
A.Cappelli, P.Valtancoli and  L.Vergnano, ``Isomonodromic Properties of the 
Seiberg-Witten Solution'',  hep-th/9710248.

\bibitem{Mgur}
A.Marshakov, ``Exact solutions to quantum field theories and integrable
equations", Mod.Phys.Lett. {\bf A11} (1996) 1169; hep-th/9602005

\bibitem{xxx}
A.Gorsky, A.Marshakov, A.Mironov and A.Morozov,
``$N=2$ supersymmetric QCD and integrable spin chains:
rational case $N_f<2N_c$", Phys.Lett. {\bf B380}
(1996) 75; hep-th/9603140

\bibitem{xyz}
A.Gorsky, A.Marshakov, A.Mironov and A.Morozov, ``A note on
spectral curve for the periodic homogeneous XYZ spin chain", hep-th/9604078

\bibitem{m}
A.Marshakov,
``From Nonperturbative Supersymmetric Quantum Gauge Theories to
Integrable Systems", in {\sl Proceedings of 10th International
Conference ``Problems of Quantum Field Theory''}, Dubna 1996,
hep-th/9607159;
``Non-\-perturbative Quantum Theories and Integrable Equations",
Int.J.Mod.Phys. A12 (1997) 1607-1650; hep-th/9610242;
``On Integrable Systems and Supersymmetric Gauge Theories",
Theor.\& Math. Phys., July 1997; hep-th/9702083

\bibitem{nikita}
N.Nekrasov,
``Five dimensional gauge theories and relativistic integrable systems",
hep-th/9609219

\bibitem{wdvv1}
A.Marshakov, A.Mironov and A.Morozov,
``WDVV-like equations in N=2 SUSY Yang-Mills Theory", Phys.Lett.
{\bf B389} (1996) 43-52; hep-th/9607109;\\
A.Marshakov, A.Mironov and A.Morozov, ``WDVV Equations from Algebra of
Forms", Mod.Phys.Lett. {\bf A12} (1997)
773-788; hep-th/9701014

\bibitem{wdvv}
A.Marshakov, A.Mironov and A.Morozov,
``More evidences for the WDVV equations in N=2 SUSY Yang-Mills theory",
hep-th/9701123;\\
A.Mironov, ``WDVV equations in Seiberg-Witten theory and associative
algebras", hep-th/9704205

\bibitem{Go96}
A. Gorsky, ``Branes and integrability in the N=2 SUSY YM theory",
hep-th/9612238

\bibitem{W} E.Witten, ``Solution of N=2 supersymmetric theories via M theory",
hep-th/9703166

\bibitem{revdon}
R.Donagi, ``Seiberg-Witten integrable systems", alg-geom/9705010

\bibitem{MMaM}
A.Marshakov, M.Martellini and A.Morozov, ``Insights and Puzzles from
Branes: 4d SUSY Yang-Mills from 6d Models", hep-th/9706050;\\
A. Marshakov, ``Seiberg-Witten Theory, Integrable Systems and D-branes",
hep-th/9709001

\bibitem{ggm}
A.Gorsky, S.Gukov and A.Mironov,
``N=2 supersymmetric field theories, integrable systems and
their stringy/brane origin -I", hep-th/9707120;\\
A.Gorsky, S.Gukov and A.Mironov,
``SUSY field theories, integrable systems and
their stringy/brane origin -II", hep-th/9710239

\bibitem{DKN} B.Dubrovin, I.Krichever and S.Novikov, in
{\sl Sovremennye problemy matematiki (VINITI), Dynamical systems - 4}
(1985) 179

\bibitem{FT} L.Faddeev and L.Takhtadjan, {\sl Hamiltonian Approach to
the Theory of Solitons}, 1986

\bibitem{Skl}
E.Sklyanin, Func.Anal \& Apps. {\bf 16} (1982) 27\\
E.Sklyanin, Func.Anal \& Apps. {\bf 17} (1983) 34

\bibitem{fumat} 
A.Hanany and Y.Oz, ``On the Quantum Moduli Space of Vacua of N=2 
Supersymmetric $SU(N_c)$ Gauge Theories", Nucl. Phys. {\bf B452} (1995) 283; 
hep-th/9505075;\\ 
P.Argyres, M.Plesser and A.Shapere, ``The Coulomb Phase of N=2 
Supersymmetric QCD", Phys.Rev.Lett. {\bf 75} (1995) 1699-1702; 
hep-th/9505100;\\ 
J.Minahan and D.Nemeschansky, ``Hyperelliptic curves for Supersymmetric 
Yang-Mills", Nucl.Phys. {\bf B464} (1996) 3-17; hep-th/9507032;\\ 
P.Argyres and A.Shapere, ``The Vacuum Structure of N=2 SuperQCD with 
Classical Gauge Groups", Nucl.Phys. {\bf B461} (1996) 437-459; 
hep-th/9509175;\\ 
A.Hanany, ``On the Quantum Moduli Space of N=2 Supersymmetric Gauge 
Theories", Nucl.Phys. {\bf B466} (1996) 85-100; hep-th/9509176 

\bibitem{theya} 
A. Brandhuber, N. Itzhaki, V. Kaplunovsky, J. Sonnenschein and 
S. Yankielowicz, 
``Comments on the M Theory Approach to N=1 SQCD and Brane Dynamics",
Phys.Lett. {\bf B410} (1997) 27-35; hep-th/9706127

\bibitem{kundu}
A.Kundu, ``Generation of a quantum integrable class of
discrete time or relativistic periodic Toda chains",
hep-th/9403001\\
S.Kharchev, ``Twisted systems of the $XXZ$ type", preprint ITEP, 1996;
unpublished

\bibitem{theisen}
A.Brandhuber, N.Itzhaki, J.Sonnenschein, S.Theisen and S.Yankielowicz,
``On the M theory approach to (compactified) $5d$ field theories",
hep-th/9709010

\bibitem{Sei}
N.Seiberg, ``Five-dimensional SUSY field theories, nontrivial fixed
points, and string dynamics", Phys.Lett. {\bf B388} 753; hep-th/9608111\\
K.Intriligator, D.R.Morrison and N.Seiberg, ``Five-dimensional
supersymmetric gauge theories and degenerations of Calabi-Yau
Spaces", Nucl.Phys. {\bf B497} (1997) 56; hep-th/9702198\\
N.Seiberg and D.Morrison, ``Extremal transitions and five dimensional
supersymmetric fields theories", Nucl.Phys. {\bf B483} (1997) 229;
hep-th/9609070

\bibitem{mamont} A.Beilinson and A.Levin, ``Elliptic polylogarithms", {\sl
Proceedings of Symposia in Pure Mathematics}, Vol.55, Part 2 (1994) 126-196\\
V.Kuznetsov, F.Nijhoff and E.Sklyanin, ``Separation of
variables for the Ruijsenaars system", hep-th/9701004\\
J.Harvey and G.Moore, ``Algebra, BPS states, and strings", 
Nucl.Phys. {\bf B463} (1996) 315-368; hep-th/9510182

\bibitem{d6}
I.Brunner and A.Karch, "Branes and six dimensional fixed points",
hep-th/9705022\\
U.H.Daniellson, G.Ferretti, J.Kalkkinen and P.Stjernberg,
"Notes on Supersymmetric Gauge Theories in Five and Six Dimensions",
Phys.Lett. {\bf B405} (1997) 265-270; hep-th/9703098\\
O.Ganor, N.Seiberg and D.Morrison
"Branes, Calabi-Yau spaces, and toroidal compactifications
of the $N=1$ six dimensional $E_{8}$ theory", Nucl.Phys. {\bf B487}
(1996) 93; hep-th/9610198\\
E.Witten, "New "gauge" theories" in six dimensions", hep-th/9710065\\
B.Kol, ``On 6d "gauge" theories with irrational theta angle", hep-th/9711017

\bibitem{nl}  N.Nekrasov and A.Lowrence,
``Instanton sums and five dimensional gauge theories",
hep-th/9706025

\bibitem{WW} E.T.Whittaker and G.N.Watson, {\sl A course of modern analysis}
(1927), 4th ed., Cambridge, 1962

\end{thebibliography}
\end{document}